\documentclass[twocolumn]{aastex631}

\usepackage{natbib}
\usepackage{tabularx}

\usepackage{enumitem}

\usepackage{graphicx}
\usepackage{url}
\usepackage{verbatim}
\usepackage{amsmath}
\usepackage{color,soul}

\begin{document}


\title{Magnetic fields in the Eos Cloud: dynamically important fields in the interface between atomic and molecular gas}

\shorttitle{B-fields in Eos}

\author[0000-0001-5996-3600]{Janik Karoly}
\affiliation{Department of Physics and Astronomy, University College London, Gower Street, London WC1E 6BT, UK}
\email{j.karoly@ucl.ac.uk} 

\author[0000-0002-8557-3582]{Kate Pattle}
\affiliation{Department of Physics and Astronomy, University College London, Gower Street, London WC1E 6BT, UK}

\author[0000-0001-5817-5944]{Blakesley Burkhart}
\affiliation{Department of Physics and Astronomy, Rutgers University, 136 Frelinghuysen Rd, Piscataway, NJ 08854, USA}
\affiliation{Center for Computational Astrophysics, Flatiron Institute, 162 Fifth Avenue, New York, NY 10010, USA}

\author[0000-0002-9583-5216]{Thavisha Dharmawardena}
\affiliation{Center for Computational Astrophysics, Flatiron Institute, 162 Fifth Avenue, New York, NY 10010, USA}
\affiliation{Center for Cosmology and Particle Physics, New York University, 726 Broadway, New York, NY 10003, USA}

\author[0000-0001-6717-0686]{B-G Andersson}
\affiliation{McDonald Observatory, University of Texas at Austin, 2515 Speedway Boulevard, Austin, TX 78712, USA}

\author[0000-0002-9593-7618]{Thomas J. Haworth}
\affiliation{Astronomy Unit, Department of Physics and Astronomy, Queen Mary University of London, Mile End Road, London, E1 4NS, UK}

\begin{abstract}

The recently-discovered Eos molecular cloud, is a CO-dark, low-density cloud located at a distance of approximately 94\,pc from the Sun which does not appear to have formed stars at any point in its history. In this paper we investigate the magnetic fields in the Eos cloud, near the interface between the atomic Cold Neutral Medium (CNM) and molecular gas, using dust emission and extinction polarimetry. A Histogram of Relative Orientation analysis shows that the magnetic field is preferentially parallel to the density structure of the cloud, while a Davis-Chandrasekhar-Fermi analysis finds magnetic field strengths of 6$\pm$3\,$\mu$G across the Eos cloud and 12$\pm$4\,$\mu$G in the somewhat denser MBM 40 sub-region. These results are consistent with a previous estimate of magnetic field strength in the Local Bubble and suggest that the fields in the Eos cloud are dynamically important compared to both gravity and turbulence. Our findings are fully consistent with the expected behavior of magnetized, non-self-gravitating gas near the CNM/molecular cloud boundary.
\end{abstract}

\section{Introduction}

Magnetic fields appear to play a crucial role in the evolution of molecular clouds, influencing processes such as star formation, turbulence, and stellar feedback interactions in the interstellar medium (ISM) of galaxies. 
\citep{10.1093/mnras/116.5.503, 1991ApJ...373..169M, 2007ARA&A..45..565M,2022preStellarReview}. 
 The newly-discovered Eos molecular cloud \citep{burkhart2025} has been identified as the one of the nearest molecular cloud to the Solar System, at a distance of 94\,pc. This low-density, relatively diffuse and non-star-forming \citep{saxena2025} molecular cloud provides an unprecedented opportunity to investigate magnetic fields in the critical intermediate regime between the cold neutral medium (CNM) and gravitationally bound sites of star formation \citep{crutcher2010}.

Zeeman H\,\textsc{i} measurements show that CNM gas is strongly magnetically subcritical (magnetic fields dominate over self-gravity), with a well-defined median magnetic field strength of $6.0\pm1.8$\,$\mu$G which is in approximate equipartition with turbulence \citep{Heiles_2005}. At higher densities, within molecular clouds, OH and CN Zeeman measurements indicate that the maximum magnetic field strength shows a power-law increase with increasing density \citep{crutcher2010,2012ARA&A..50...29C}. The transition from magnetic field strength being independent of gas density in the CNM to a power-law increase with density in molecular gas has been placed at a volume density $n_{\rm H}\approx 300$\,cm$^{-3}$ \citep{crutcher2010}, and at a column density of $N_{\rm H}\sim 10^{22}$\,cm$^{-2}$ \citep{2019CrutcherKemball}. While the exact density at which this transition occurs remains under debate \citep[e.g.,][and refs. therein]{2022preStellarReview}, it is generally agreed that the transition itself is related to the onset of gravitational collapse, which under flux freezing conditions will increase the magnetic field strength of a cloud \citep{mestel1966}.

Dust polarization observations from the \textit{Planck} Observatory \citep{2016A&A...596A.103P}, \textit{Herschel}, BLASTpol \citep{2013A&A...550A..38P,soler2017} and the James Clerk Maxwell Telescope \citep[e.g.][]{kwon2022} have revealed a strong connection between filamentary structures and magnetic fields in star-forming regions. These studies show a transition in the relationship between magnetic field and column density structure in star-forming clouds, with low-density cloud material preferentially running parallel to the magnetic field orientation in the cloud, and denser, star-forming, filaments typically oriented perpendicular to the magnetic field orientation \citep{2013A&A...550A..38P,2013ApJ...774..128S,2017A&A...607A...2S,2020MNRAS.496.4546H,2021MNRAS.503.5425B}. This shift in alignment is likely driven by gravitational instability and/or the convergence of gas flows \citep{2017A&A...607A...2S}, and typically occurs at a column density $N_{\rm H}\sim 10^{21.7}$\,cm$^{-2}$ \citep{Planck2016_xxxv}, similar to that of the transition in the relationship between magnetic field strength and gas density identified by \citet{crutcher2010}. However, the universality of this transition column density remains poorly constrained \citep{2022preStellarReview}, as does its relationship, if any, with the \citet{crutcher2010} transition. Studies of magnetic fields in the CNM/molecular cloud transition, and in non- or marginally self-gravitating molecular clouds, are therefore crucial in order to constrain the strength and dynamic importance of magnetic fields in this critical density regime.

In this work, we present an in-depth study of the magnetic field structure in the Eos cloud, a diffuse, almost entirely CO-dark \citep{burkhart2025}, non-star-forming \citep{saxena2025} molecular cloud, using dust emission and extinction polarimetry. The Eos cloud is positioned at the high-latitude edge of the North Polar Spur (Loop I), a prominent X-ray and radio structure. The nature of the observed far-ultraviolet (FUV) H$_2$ fluorescent emission in Eos suggests that it is interacting with Loop I, with X-ray emitting hot gas contributing to the excitation of the molecules. The sole site of CO emission in Eos corresponds with the high-latitude cloud MBM 40 \citep{magnani1985}. Distance estimates for MBM 40 vary, and could place it within or behind the Eos molecular cloud \citep{sun2021}.
These unique characteristics place Eos in a crucial astrophysical context, providing an opportunity to study magnetic fields in a CO-dark molecular cloud at the interface of both a hot-cold boundary layer and the atomic-to-molecular boundary layer. 

This paper is organized as follows: In Section \ref{sec:data}, we provide an overview of the starlight polarization data, Planck polarization maps, and Eos cloud datasets used in this study. Section \ref{sec:results}
describes our results and analysis. We discuss our results in Section \ref{sec:dis}. Section \ref{sec:con} presents our conclusions.

\section{Data}
\label{sec:data}

\begin{figure*}[!t]
  \centering
  \includegraphics[width=0.8\textwidth]{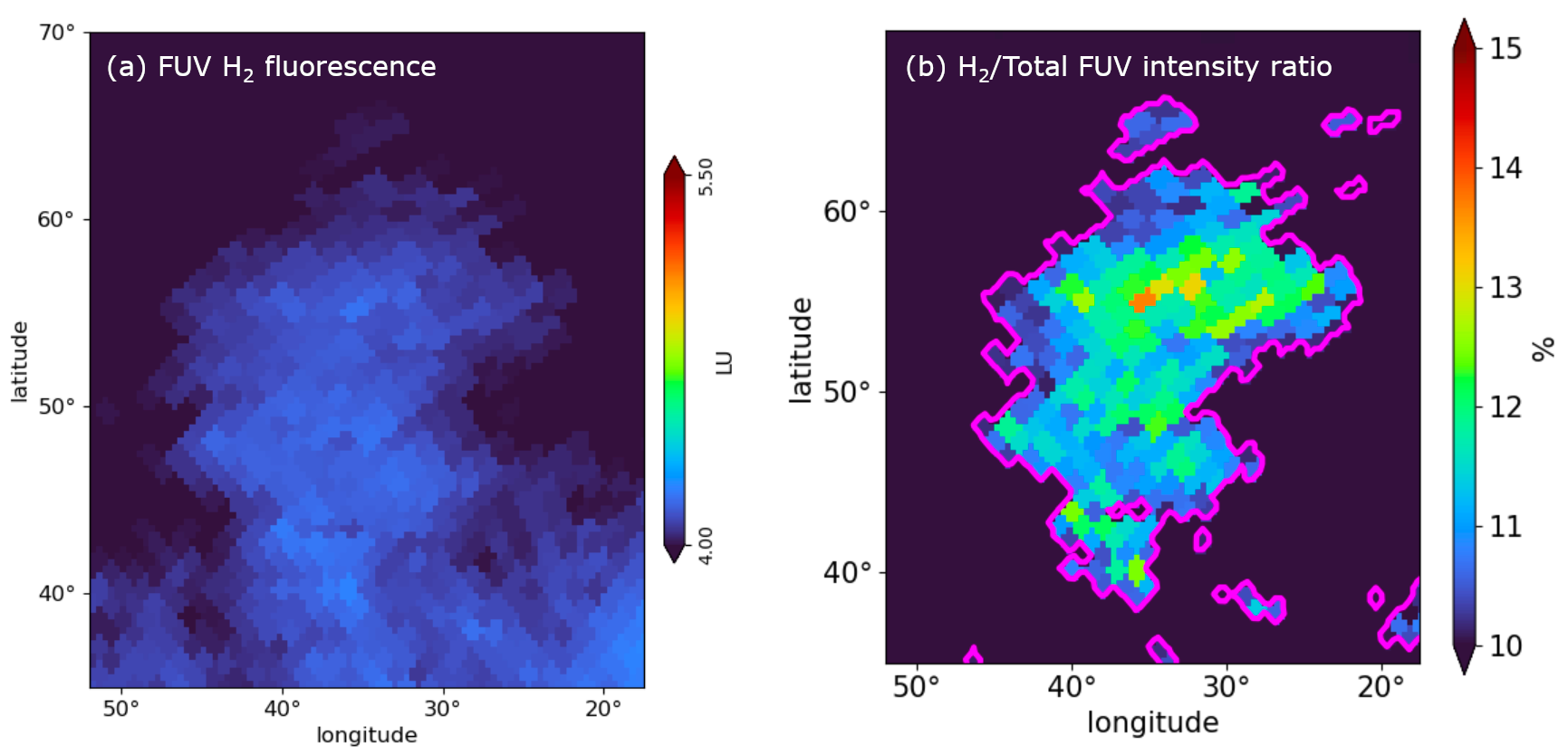}
  \caption{The Eos cloud. \textit{Left:} FIMS/SPEAR FUV H$_2$ fluorescence emission map, in log line units (LU: photons cm$^{-2}$ sr$^{-1}$ s$^{-1}$), first presented by \cite{jo2017ApJS..231...21J}. \textit{Right:} Map of the ratio of H$_2$ intensity to total FUV intensity, in percentage points (\%). The location of the Eos cloud is outlined and appears as a bright fluorescent feature. The on-sky boundary for the Eos cloud is based on the magenta contours, which outline the 10\% ratio of H$_{2}$ intensity.}
  \label{fig:eos} 
\end{figure*}

The FUV H$_2$ observations used to identify and define the Eos cloud are shown in Figure~\ref{fig:eos}; for a complete discussion of these observations, see \citet{burkhart2025}. We use the \citet{burkhart2025} H$_2$/total FUV ratio map, shown in Figure~\ref{fig:eos}b, to define the Eos cloud boundary throughout this work.

\subsection{Dust emission polarization data}

We retrieved 353\,GHz \textit{Planck} Commander thermal dust emission maps \citep{2015PlanckXIX} from the \textit{Planck} Legacy Archive\footnote{\url{https://pla.esac.esa.int/}} and reprojected from healpix to Cartesian coordinates on a 5$\arcmin$ (the FWHM resolution) grid. The Commander maps are made from Planck software code implementing Bayesian parametric component separation. We use the Stokes $Q$ and $U$ maps to calculate the polarization angle,

\begin{equation}
{\theta = \frac{1}{2}\arctan\left(\frac{-U}{Q}\right)} \, ,
\label{eq:theta}
\end{equation}

where $U$ and $Q$ are the measured \textit{Planck} Stokes parameters (note the negative {\rm sign on $U$, required to convert the \text{Planck} measurements to IAU convention). We calculate polarization percentage $p$,

\begin{equation}
p = 100\% \times \frac{\sqrt{Q^{2} + U^{2}}}{I} \, ,
\label{eq:p}
\end{equation}
\noindent 

where \textit{I} is total intensity (Stokes \textit{I})}. 

The plane-of-sky (POS) orientation of the magnetic field is inferred by rotating the polarization angles (eq. \ref{eq:theta}) by 90$^{\circ}$, assuming that the polarization arises from elongated dust grains aligned perpendicular to the magnetic field \citep{2015Andersson}.

\begin{figure*}[!t]
  \centering
  \includegraphics[width=0.8\textwidth]{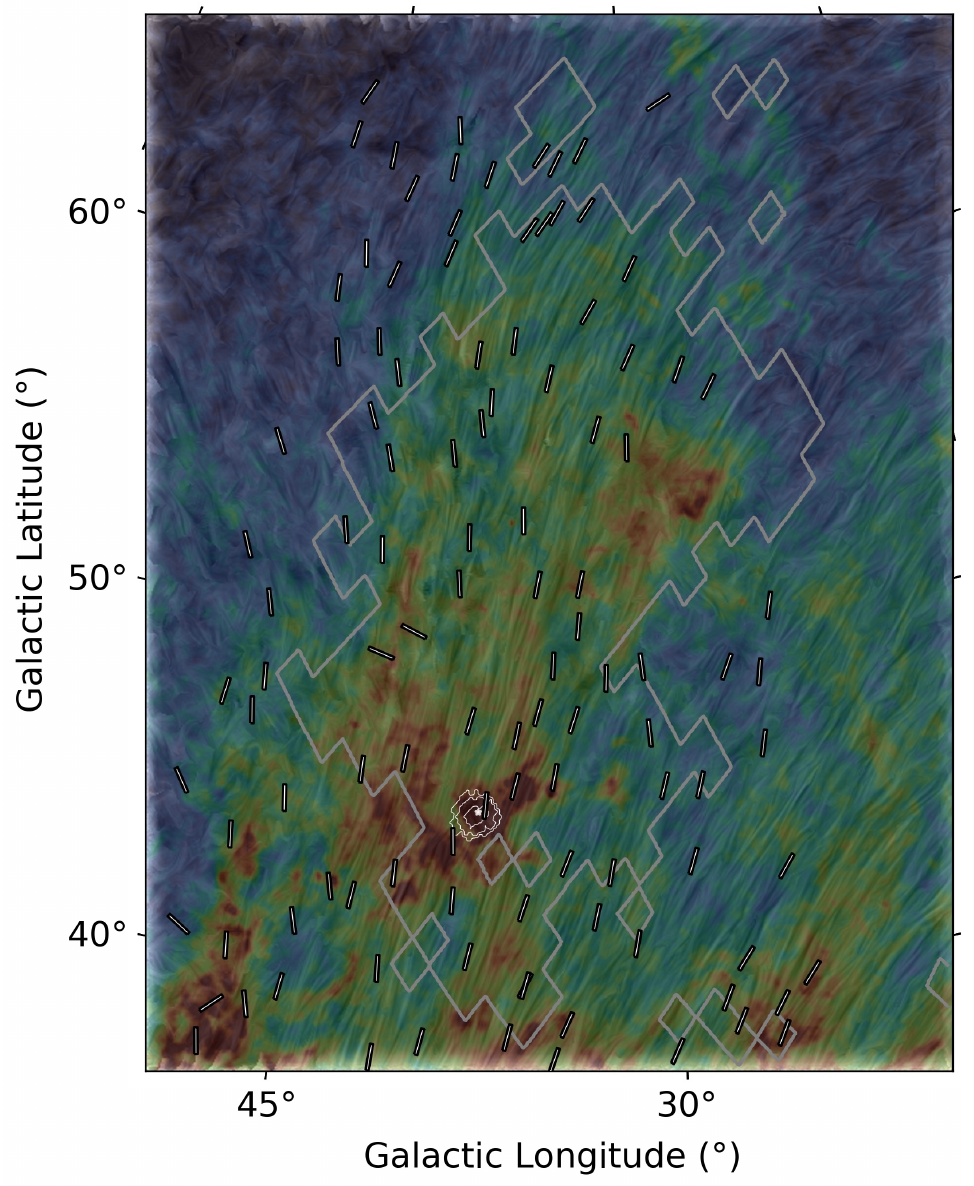}
  \caption{Magnetic fields in the Eos cloud. The image shows Stokes \textit{I} dust emission observed by \textit{Planck} at 353\,GHz, with overlaid texture showing a line integral convolution \citep[LIC;][]{lic_cabral} visualization of the magnetic field observed by \textit{Planck}. Optical vectors are overlaid as white lines. The gray contour marks the bounds of the Eos cloud as defined using FUV H$_2$ observations \citep{burkhart2025}. The white contours are Planck CO (1-0) observations which show the location of MBM 40 and the star shows the central coordinates of MBM 40 from \citet{2022ApJS..262....5D}.}
  \label{fig:lic} 
\end{figure*}

\subsection{Dust extinction (starlight) polarization data}

We extracted dust extinction (starlight) polarization fractions and angles for the Eos cloud region from {\rm the catalogue presented by} \citet{berdyugin2014}. These vectors are all plotted in Figure~\ref{fig:lic}.

If we limit the selection to vectors only spatially coincident with the Eos molecular cloud as determined by the grey contours in Figure~\ref{fig:lic}, we obtain a sample of 51 vectors.
The polarization measurements have an average signal-to-noise ratio of 13 with a minimum of 3.5 (only two targets have S/N$\leq$5) and a maximum of 34.0.

Based on Gaia DR3 parallaxes \citep{gaiadr3}, available for only 25 targets, the target stars are located at distances of between 145 and 500 pc from the Sun, with most, 17, being $<$300\,pc. 
 This places the full sample near but behind the Eos cloud \citep{burkhart2025}.

For starlight (optical) polarization, the polarization vector orientation directly traces the POS magnetic field orientation \citep{2015Andersson}.

\subsection{21-cm H\,{\sc i} emission}

The Galactic Arecibo L-Band Feed Array H\,{\sc i} (GALFA-H\,{\sc i}) Survey is a comprehensive project aimed at mapping the neutral hydrogen in the Milky Way at a spatial resolution of 4 arcmin and a spectral resolution of 0.18 $\rm km \; s^{-1}$ \citep{2011ApJS..194...20P}. H\,{\sc i} is expected to be observed in and around molecular clouds as it is required for shielding molecules from dissociation from background UV photons \citep{2012ApJ...748...75L,2016ApJ...829..102I,2018ApJ...856..136P}. In deeper regions of the cloud, the H$_2$ can self-shield against dissociating UV radiation, but the outer layers will still have substantial amounts of H\,{\sc i} due to ongoing photodissociation \cite{2015ApJ...811L..28B,Sternberg_2021}. The fluorescent FUV emission traces the atomic-to-molecular cloud boundary (i.e., the H$_2$ to H\,{\sc i} transition), and indeed, an excellent agreement is observed between the 21-cm map and the FIMS/SPEAR contours \citep{burkhart2025}.  We use the GALFA-H\,{\sc i} Survey data in our analysis of magnetic field strength, as described in Section~\ref{sec:results_dcf}, below.

\section{Results}
\label{sec:results}

Figure~\ref{fig:lic} displays the magnetic field inferred from \textit{Planck} observations, shown using the line integral convolution \citep[LIC;][]{lic_cabral} representation. A clear north--south magnetic field orientation can be seen, which aligns with the orientation of the cloud, as defined using both the FUV H$_{2}$ boundary contour and the Stokes \textit{I} emission. Starlight polarization vectors from \citet{berdyugin2014}, which directly trace the magnetic field orientation, are shown as white vectors\footnote{These, and other, polarization segments shown in this work should formally be referred to as `half-vectors' or `pseudovectors' since polarization angles are defined over the range $0-180^{\circ}$. We use the shorthand `vector' in keeping with general practice in the field.}.

The magnetic field position angles of the starlight polarization vectors and the \textit{Planck} vector along the same line-of-sight (LOS) are plotted in the left panel of Figure~\ref{fig:pvsi}. The vectors show good agreement, with two of the noticeable outliers corresponding to the two approximately east/west-aligned vectors in the center left of Figure~\ref{fig:lic}. The Planck magnetic field in the location of these two vectors does look more disturbed (see Figure~\ref{fig:lic}) and it is a region of lower H$_2$/FUV ratio (see Figure~\ref{fig:eos}), so perhaps the magnetic field is disrupted by local turbulence or a shock. In addition, while most stars are not strongly inherently polarized, with only single band optical data we cannot exclude the possibility that the star is intrinsically polarized. Since the stars that we consider are located behind the Eos cloud and so their polarization is induced by the dust layer of the Eos molecular cloud, agreement with \textit{Planck} suggests that we are successfully tracing the magnetic field of the molecular cloud with the \textit{Planck} observations. In addition, since this is a high latitude cloud, we do not expect significant foreground or background contamination of the \textit{Planck} emission.

In this work we use 353\,GHz Stokes $I$ intensity as a proxy for column density. A direct proportionality between intensity and column density strictly holds only for isothermal dust emission \citep{hildebrand1983}, but since the Eos cloud is of limited total column density, and lacks significant sources of internal heating \citep{burkhart2025,saxena2025}, it is a reasonable approximation in this environment.

\subsection{Polarization Properties of Eos}
\label{subsec:pol}

We plot polarization efficiency (p/A$_{0}$) versus Gaia-derived A$_{0}$ values \citep{gaiadr3} for the optical polarization vectors in the center panel of Figure ~\ref{fig:pvsi}. The expected relationship between these quantities is a power law, with the power index indicating how efficient the dust grain alignment is \citep{2015Andersson}. Similarly, but in the submillimeter (dust emission) regime, we plot non-debiased polarization percentage (see eq.\,\ref{eq:p}) versus Stokes \textit{I} in the right panel of Figure~\ref{fig:pvsi}. We fitted the data by creating a logarithimically-spaced 2D histogram of Stokes \textit{I} and $p_{frac}$, and for each binned value of Stokes \textit{I}, evaluated the mean value of $p_{frac}$ over bins containing at least 1\% of the total number of pixels ($\approx 140$ pixels).  We then fitted these mean values, shown as black diamonds on the upper right panel of Figure~\ref{fig:pvsi}, with the power law function. Again, we expect to see a power law relation \citep{2015Andersson,katerice}. 

In both cases, a power index of 0 indicates perfect grain alignment while a value of -1 indicates complete loss of grain alignment. In addition, a power-law index of -0.5 is expected for constant grain alignment efficiency in a turbulent medium \citep{jones89}. For the optical polarization observations, the best-fit power index is $-0.61\pm0.08$ and for the \textit{Planck} polarization observations, the power index is $-0.51\pm0.03$. In both cases, this suggests that efficient grain alignment is maintained throughout Eos and that the magnetic field is well traced by dust polarization in both the optical and submillimeter regimes.

The fit for the optical polarization plot (middle panel of Figure~\ref{fig:pvsi}) does not fit the data well at higher A$_{0}$ values and the overall fit quality is quantified by a coefficient of determination R$^2$=0.56, indicating that our model can explain slightly over half of the variation in the data. We note the single power-law fit is not a complete model and can miss transitions towards less efficient alignment at higher A$_{0}$ \citep{2015Andersson}. Rather it is used as a diagnostic of the overall efficiency of the alignment and coupled with the good agreement between magnetic fields traced by optical and submillimeter polarization, we can assume there is good alignment along the line of sight. The generally accepted mechanism enabling this would be Radiative Alignment Torques \citep[RATs,][]{1976Ap&SS..43..291D,hoang08,2015Andersson}. RAT requires light with a wavelength ($\lambda$) less than the grain diameter (d$_{\rm grain}$) and if we assume the upper end of the \citet{mrn77} grain distribution, then d$_{\rm grain}\sim$0.5\,$\mu$m and so $\lambda<$0.5\,$\mu$m is needed (just short of V-band) where extinction would only be A$_{\rm V}\sim$1.15 mag, still allowing for plenty of radiation. Newer grain size estimates also have exponential tails beyond 0.25\,$\mu$m \citep[eg.][]{clayton2003} so RAT alignment is an expected mechanism in this cloud.

We also note there are high polarization efficiency values (p/A$_{0}>3\%/$mag) shown in this plot. This could be dependent on the composition of grains \citep[such as super-paramagnetic inclusions,][]{lazarian08,thiem16}. Grains with super-paramagnetic inclusions (MRATs) may be necessary if the large p/A$_{0}$ at low A$_{0}$ are true, but we also note that the large p/A$_{0}$ values originate from lower A$_{0}$ values and so are subject to systematic errors. Determining if grain alignment in Eos is due to MRATs versus RATs is not answerable in this study and would require full spectro-polarimetry follow-up of those lines of sight.

\subsubsection{Planck Polarization Angle Dispersion Function}

To further test the efficiency of grain alignment in the Eos cloud, we investigated the relationship between p$_{frac}=p/100$ and the polarization angle dispersion function, $\mathcal{S}(x, \delta)$ (henceforth $\mathcal{S}$), given by
\begin{equation}
S(x, \delta) = \left( \frac{1}{N} \sum_{i=1}^{N} (\Delta \psi_{xi})^2 \right)^{1/2} \, ,
\end{equation}
\citep{2015PlanckXIX}, where $\delta$ is a `lag' which defines an annulus of $\delta$/2 to 3$\delta$/2 around the central pixel $x$ over which $\Delta \psi_{xi}$ is summed, and
\begin{equation}
\Delta \psi_{xi} = \frac{1}{2} \arctan \left( \frac{Q_i U_x - Q_x U_i}{Q_i Q_x + U_i U_x} \right) \, .
\end{equation} 
$\mathcal{S}$ measures the spread in polarization angle orientations, and hence magnetic field directions, over the area defined by $\delta$.  We take $\delta=5\arcmin$ to match the resolution of the \textit{Planck} data. Where the magnetic field is well-ordered, $\mathcal{S}$ is small, while in disordered regions $\mathcal{S}$ increases towards a maximum of $\approx52^{\circ}$ \citep{2015PlanckXIX}. We find a majority of $\mathcal{S}$ values are $<$10$^{\circ}$ which may indicate that the large-scale magnetic field is preferentially orientated in the plane of the sky \citep[][values $>20^{\circ}$ indicate an orientation along the line-of-sight]{planckxx}. \citet{2015PlanckXIX,planckxx,planckxii} found $\mathcal{S}$ to be anti-correlated with $p_{frac}$, which they interpret to indicate that tangling of the magnetic field, rather than decreased grain alignment efficiency, leads to depolarization in dense regions within clouds.

\begin{figure*}[!t]
  \centering
   \includegraphics[width=0.3\textwidth]{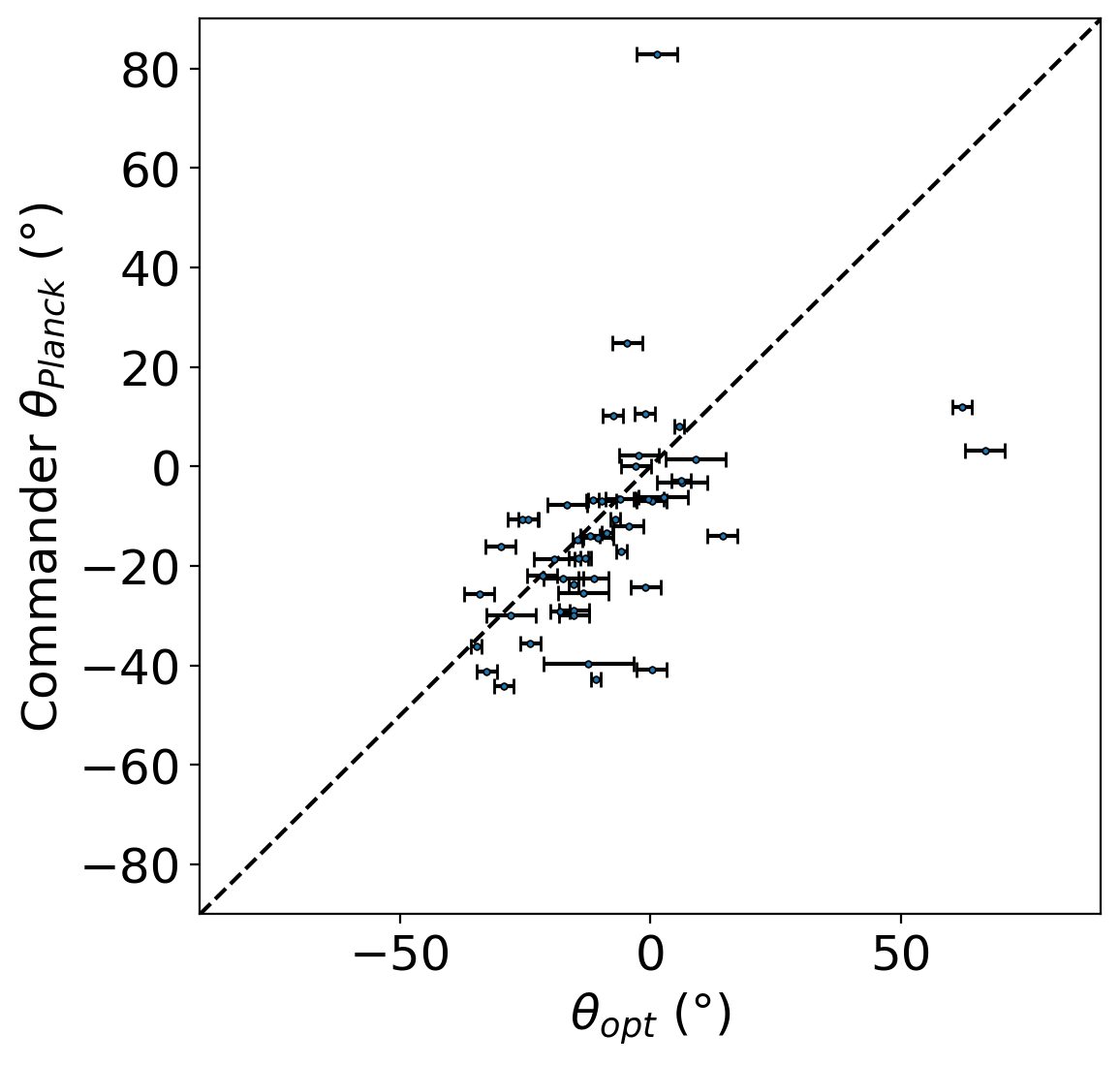}
  \includegraphics[width=0.3\textwidth]{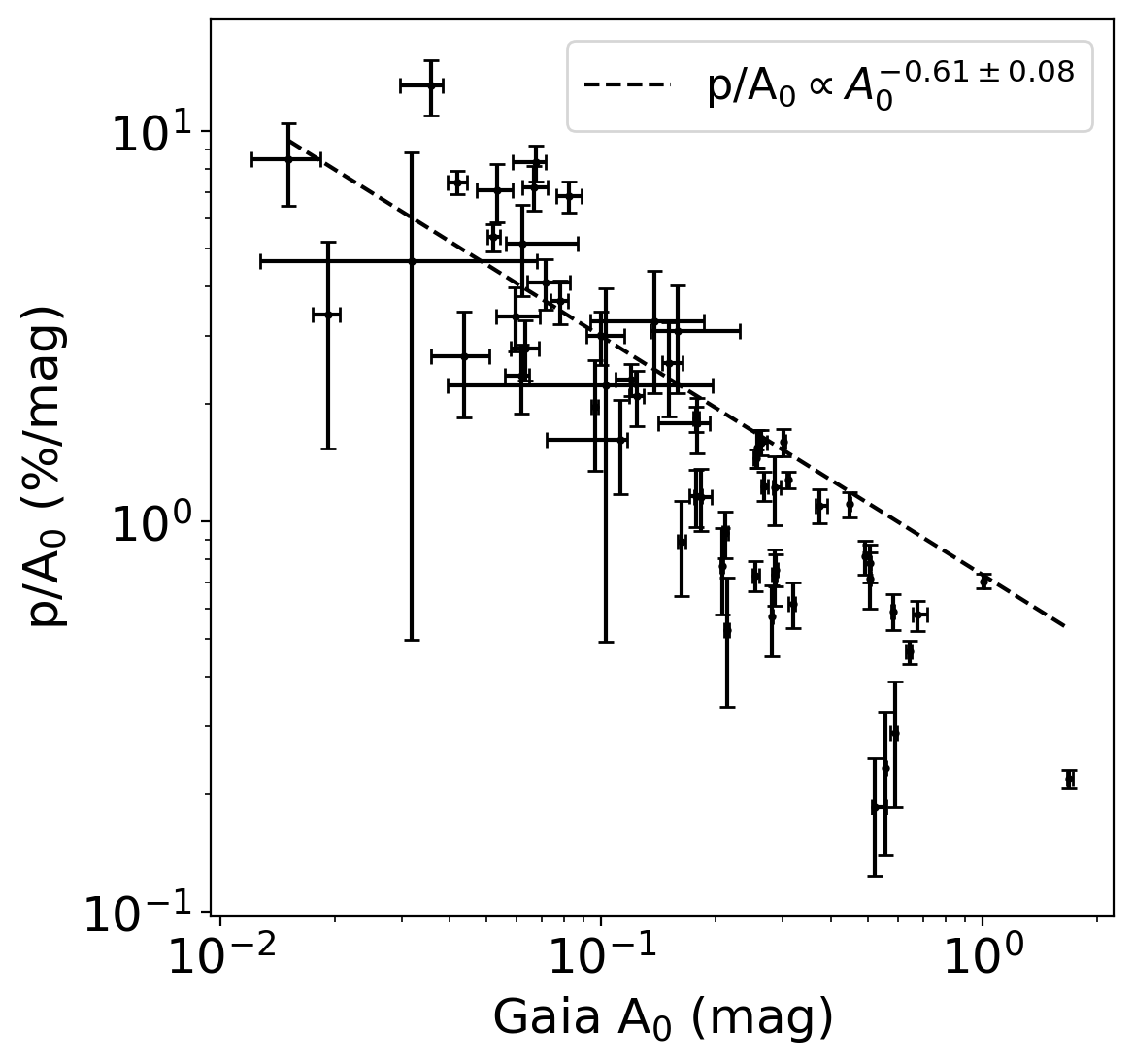}
  \includegraphics[width=0.33\textwidth]{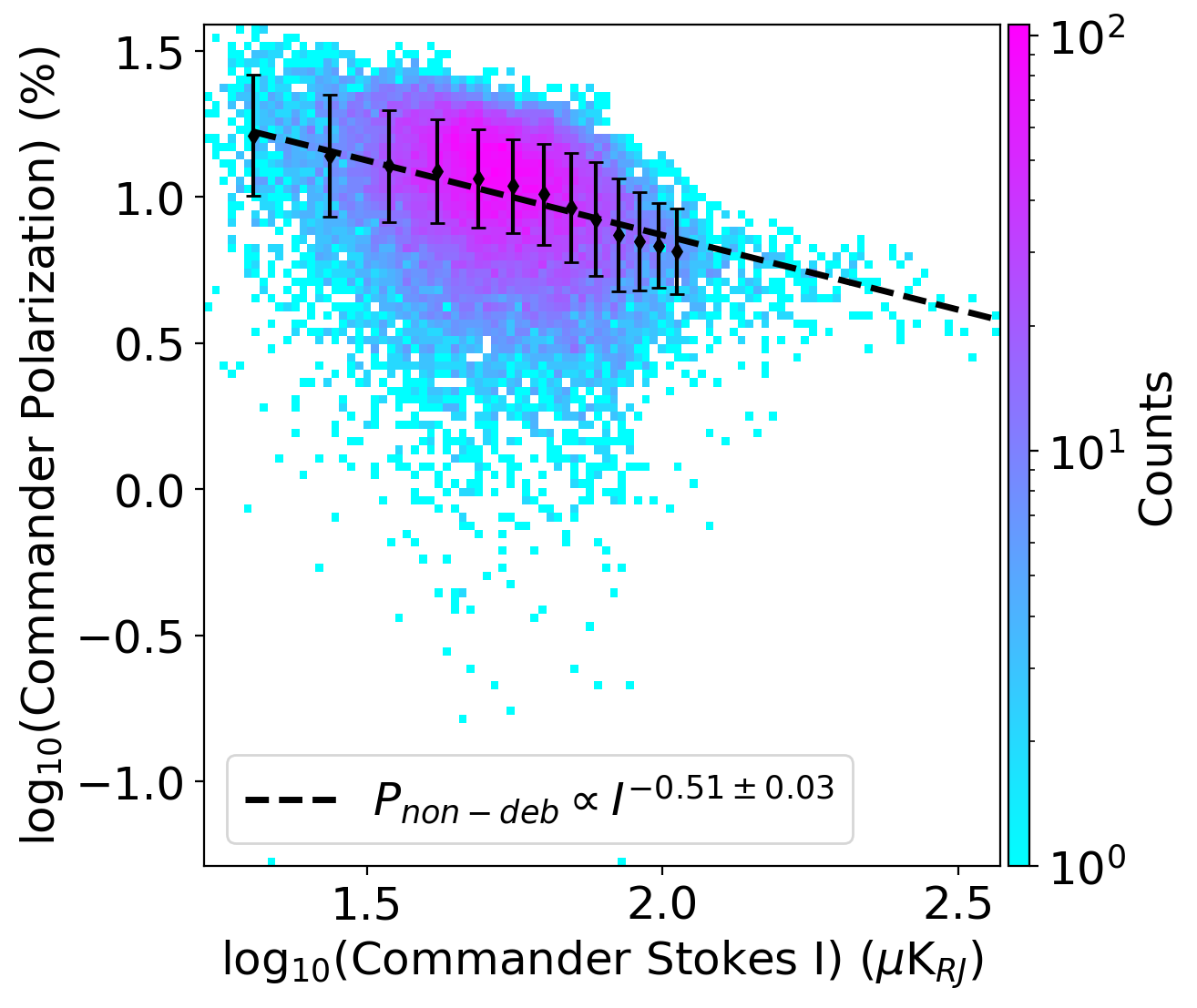} \\  \includegraphics[width=0.45\textwidth]{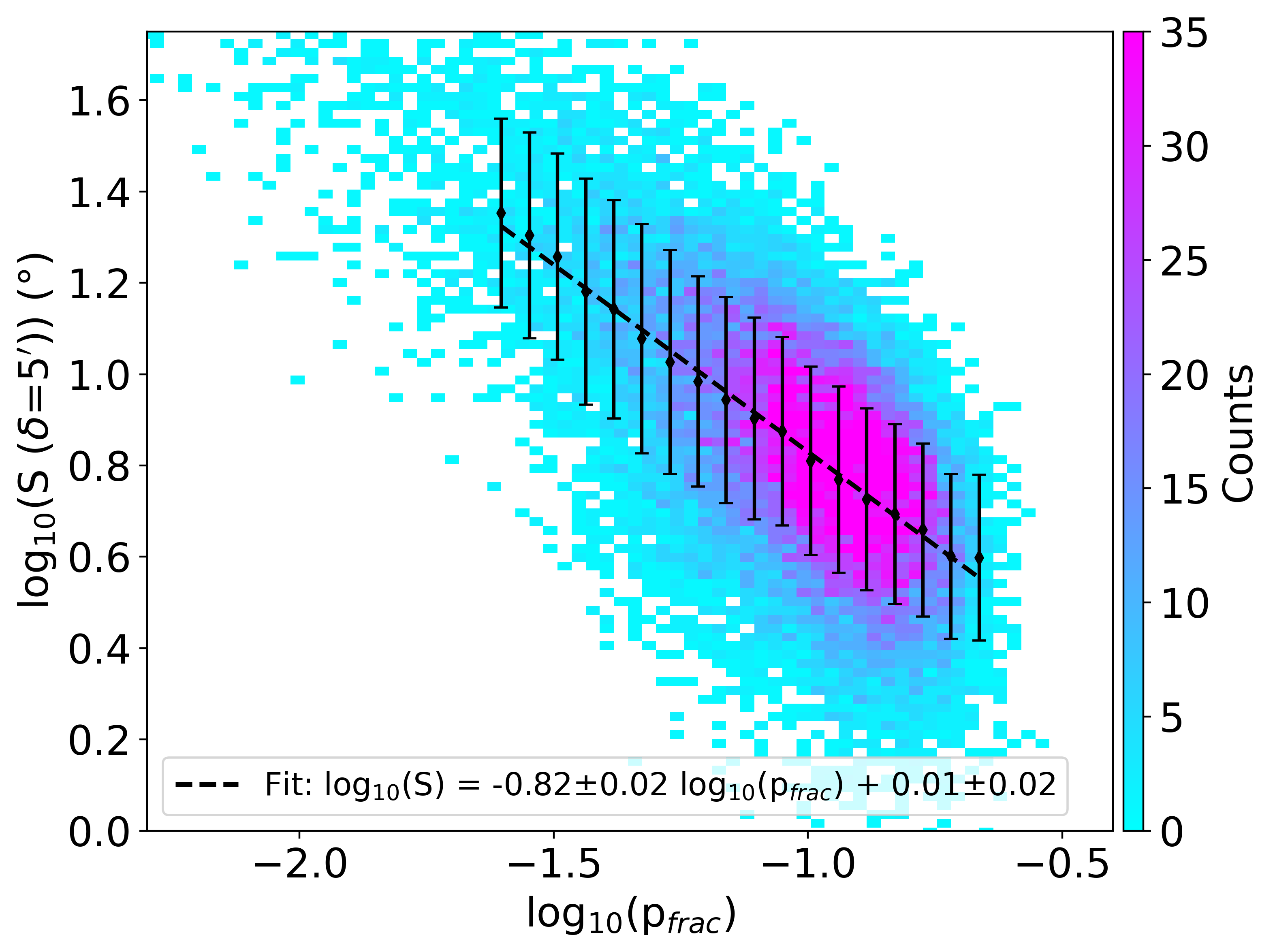} 
  \includegraphics[width=0.45\textwidth]{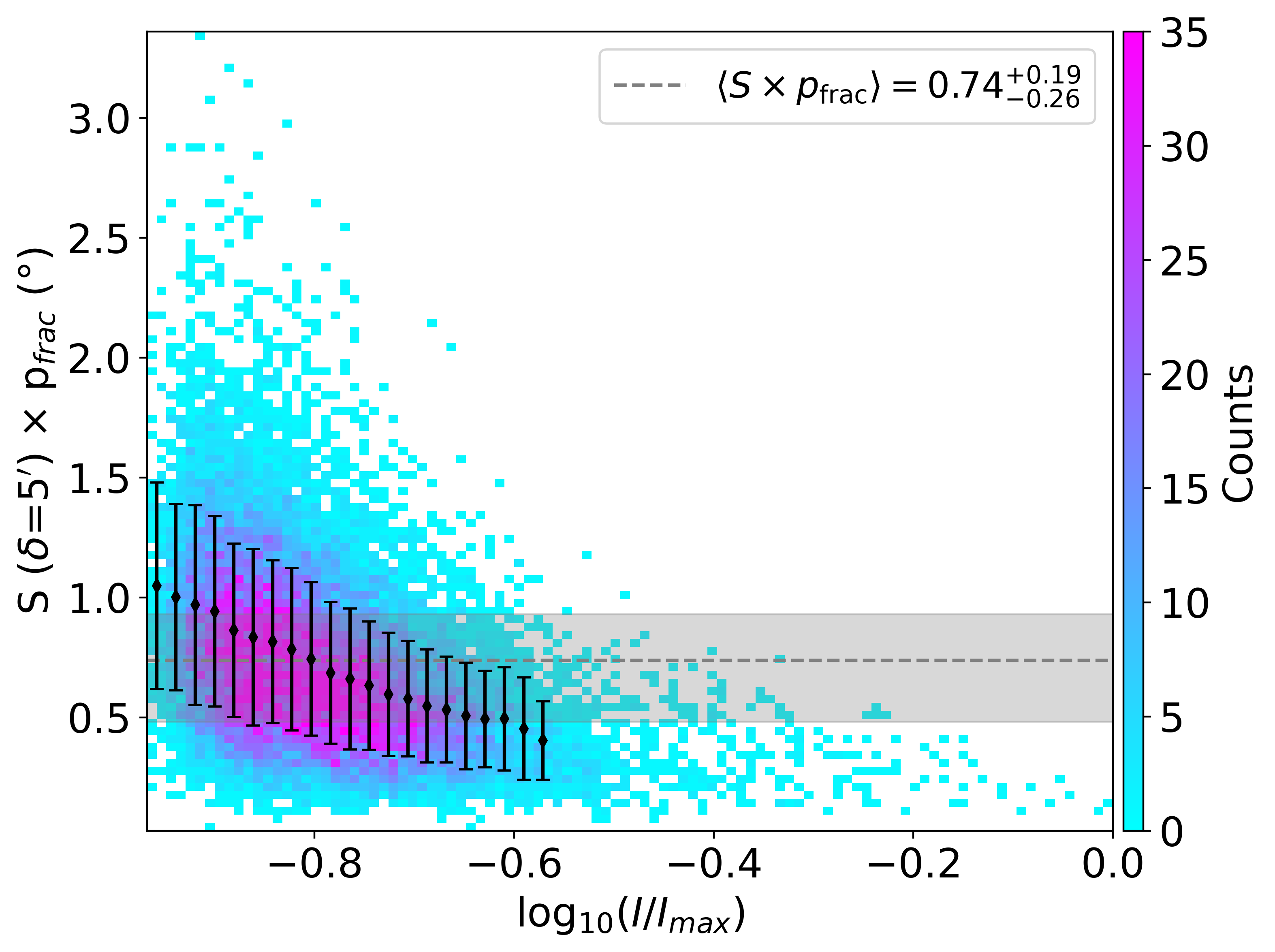} 
  \caption{\textit{Upper Left:} Optical polarization angle versus submillimeter Planck magnetic field angle (polarization angle rotated by 90$^{\circ}$). A one-to-one dashed line is plotted. Most of the position angles agree well. \textit{Upper Center:} Polarization efficiency (p/A$_{0}$) versus A$_{0}$ for the optical polarization vectors. A power law fit gives an index of $-0.61$, indicating good grain alignment \citep{2015Andersson}. \textit{Upper Right:} Non-debiased polarization percentage (see Eq.\,\ref{eq:p}) versus Stokes \textit{I}. A power-law fit gives an index of $-0.51$, again indicating good grain alignment. \textit{Lower Left:} A plot of the polarization structure function $\mathcal{S}$ evaluated with a lag of 5$\arcmin$ versus non-debiased polarization fraction (p$_{frac}$). A simple linear fit of log(S) versus log(p$_{frac}$) \citep[see][]{2015PlanckXIX,planckxx,valentin2020} is shown as a black dashed line. More details are given in the text. \textit{Lower Right:} A plot of S$\times$p$_{frac}$ versus normalized intensity which is used as a proxy for column density. A slight decreasing trend can be seen. The mean S$\times$p$_{frac}$ value is plotted as a grey dashed line with a shaded area representing the 25\% and 75\% quartiles.}
  \label{fig:pvsi} 
\end{figure*}

$\mathcal{S}$ and p$_{frac}$ are anti-correlated in the Eos cloud, as shown in the lower left panel of Figure~\ref{fig:pvsi}. Without provided uncertainties on the \textit{Planck} Commander $Q$ and $U$ maps, we were unable to debias our polarization fraction or apply signal-to-noise cuts and so we cut the data, keeping polarization data with \textit{I}$>$40$\mu$K$_{\rm RJ}$. We fitted the data with the same method (a logarithimically-spaced 2D histogram) as earlier in Section~\ref{subsec:pol}, but over bins containing here at least 0.5\% of the total number of pixels ($\approx 100$ pixels).  We then fitted these mean values, shown as black diamonds on the lower left panel of Figure~\ref{fig:pvsi}, with a linear function \citep[cf.][]{2015PlanckXIX,planckxx,valentin2020}.  Our best-fit relation is shown as a black dashed line in the lower left panel of Figure~\ref{fig:pvsi}. Our power-law index of $-0.82\pm 0.02$ is comparable to the \citet{2015PlanckXIX} and \citet{planckxx} values of $-0.83$ and $-0.75$ respectively.

\citet{planckxx} and \citet{planckxii} showed $\mathcal{S}\propto p_{frac}^{-1}$ for clouds containing a large number of turbulent layers, while \citet{valentin2020} showed that for a single turbulent layer, $\mathcal{S}\propto p_{frac}^{-0.5}$.  Our fitted value falls between these two extremes, and is consistent with previous \textit{Planck} results, suggesting the Eos cloud contains a small number of turbulent layers along the line of sight, as might be expected for a sheetlike cloud.  The power-law index that we measure further suggests that tangled fields in the Eos cloud may contribute to the depolarization seen at high Stokes $I$, but may not be solely responsible for it \citep{2015PlanckXIX,planckxii}.

The product $\mathcal{S}\times p_{frac}$ has been suggested as a proxy for grain alignment efficiency \citep[][and refs. therein]{valentin2020}. The lower right panel of Figure~\ref{fig:pvsi} shows $\mathcal{S}\times p_{frac}$ plotted as a function of normalized intensity (a proxy for column density) for the Eos cloud.  We see a slight anti-correlation, with a mean value of 0.74$^{\circ}$. This suggests that the grain alignment efficiency may depend on local conditions such as gas density or temperature in the Eos cloud \citep{valentin2020}.  However, \textit{Planck} Commander $Q$ and $U$ maps are not provided with uncertainties with which to select vectors or debias $p_{frac}$, potentially leading to larger values of $\mathcal{S}\times p_{frac}$ at low Stokes $I$.  Therefore we are unable to state conclusively that there is a significant decrease of $\mathcal{S}\times p_{frac}$ with increasing intensity.

\subsection{Magnetic Field and Cloud Structure}
\label{subsec:hro}

 \textit{Planck} and \textit{Herschel} observations have revealed a general trend for magnetic fields to transition from being aligned parallel to column density structures in diffuse gas to perpendicular to them at high densities \citep{Planck2016_xxxv}. This transition is usually quantified using the histogram of relative orientation \citep[HRO;][]{2013ApJ...774..128S,soler2017}.  The HRO measures the distribution of angles between contours of iso-(column) density (with position angle $\psi$) and magnetic field orientation.
$\psi$ is perpendicular to the column density gradient direction in the POS.

The difference in angle between the magnetic field vector and the contour orientation vector is then given by $\phi$, termed the angle of relative orientation \citep{2013ApJ...774..128S}, in the range $|\phi|<90^{\circ}$. The distribution of $\phi$ is split into three ranges, $|\phi|<22.5^{\circ}$, $|\phi|>67.5^{\circ}$, and $22.5^{\circ}<|\phi|<67.5^{\circ}$, which represent preferentially parallel alignment, preferentially perpendicular and no preferred orientation, respectively. 

To determine how the alignment of the magnetic field with cloud structure varies with density, the map is divided into a series of column density or, as in this case, intensity bins. For each bin, a histogram of $\phi$ values is constructed. The shape parameter $\xi$ is then given by
\begin{equation}
  \xi = \frac{A_{\rm c}-A_{\rm e}}{A_{\rm c}+A_{\rm e}} \, \, ,
  \label{eq:shape}
\end{equation}
\noindent
where $A_{\rm c}$ is the area of the histogram where $|\phi|<22.5^{\circ}$ (preferentially parallel) and $A_{\rm e}$ is the area where $|\phi|>67.5^{\circ}$ (preferentially perpendicular). If $\xi<0$ there is a preferentially perpendicular alignment between the density structure and the magnetic field (i.e., density gradients are parallel to the magnetic field) and if $\xi>0$ there is preferentially parallel alignment between the magnetic field orientation and the density structure (i.e., density gradients are then perpendicular to the magnetic field).

\begin{figure}[!t]
  \centering
  \includegraphics[width=\columnwidth]{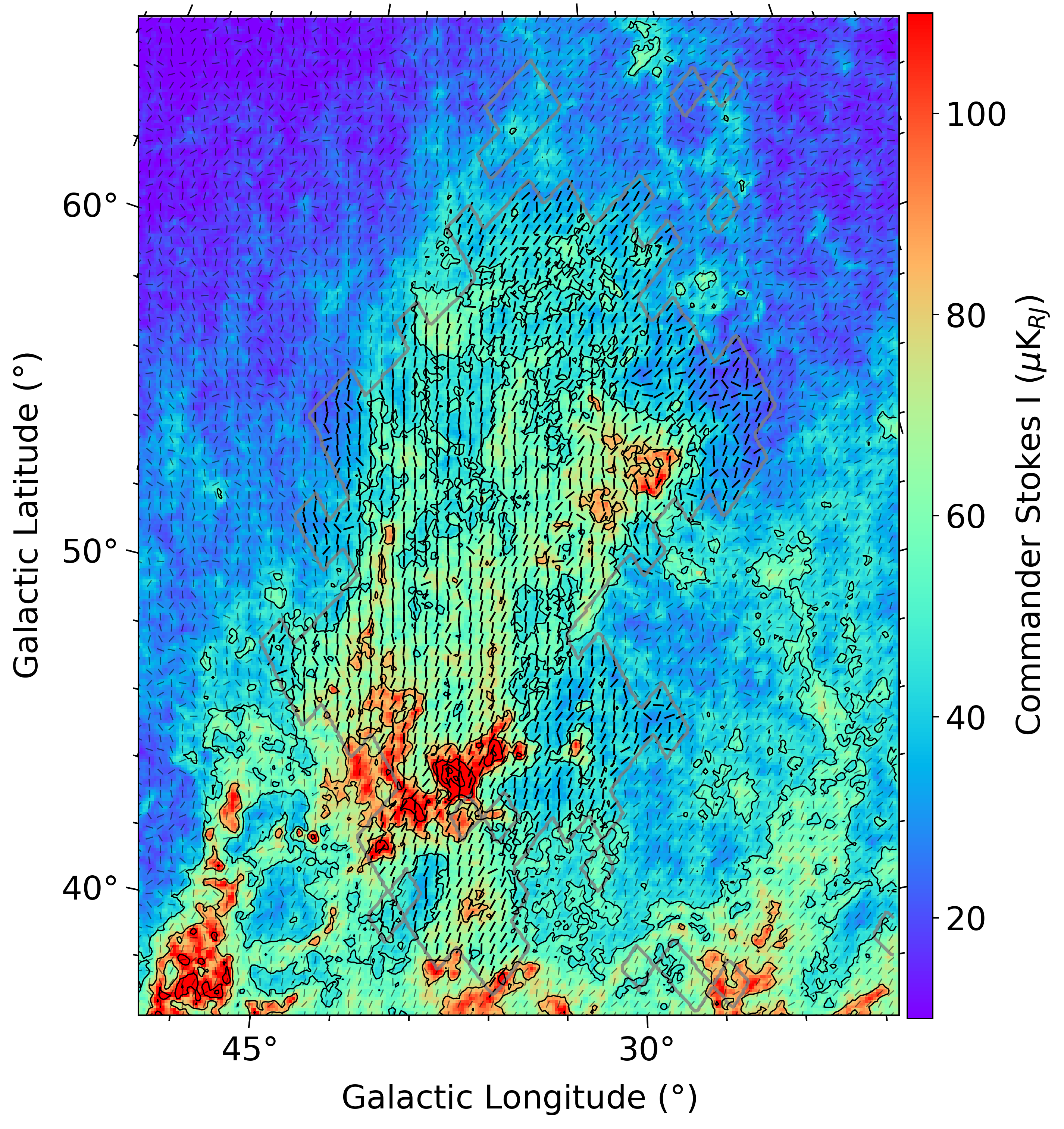} \\
  \includegraphics[width=\columnwidth]{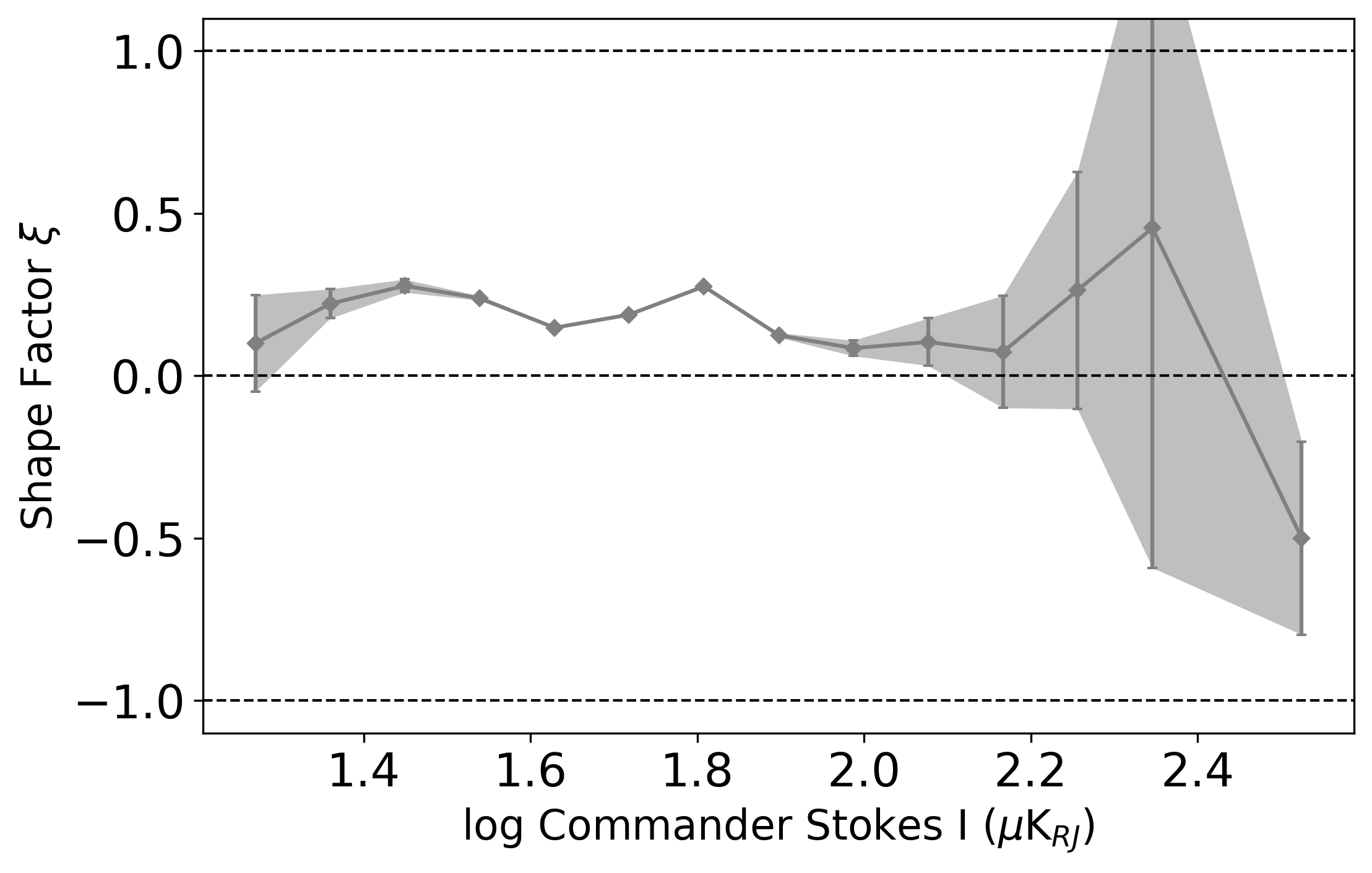}
  \caption{An assessment of the relationship between magnetic field and intensity structure (as a proxy for column density) in the Eos cloud. \textit{Upper:} \textit{Planck} magnetic field vectors plotted on the \textit{Planck} Stokes \textit{I} map. Vectors used in the HRO analysis are shown in black and bold. Other vectors are plotted more faintly. Contours of Stokes $I$ are plotted at intensities of 50, 80, 120, 200, and 300 $\mu$K$_{\rm RJ}$. \textit{Lower:} HRO \citep{2013ApJ...774..128S,soler2017} of the selected vectors in the upper panel, showing a clear tendency for magnetic fields to run parallel to intensity structure.}
  \label{fig:hro} 
\end{figure}

The upper panel of Figure~\ref{fig:hro} shows the \textit{Planck} magnetic field vectors overlaid on the \textit{Planck} dust emission. The vectors and dust emission within the cloud boundary, identified by the gray UV contour, are used to construct the HRO which is shown in the lower panel. The HRO shows a preferentially parallel orientation ($\xi > 0$) between the cloud structure and the magnetic field across almost all of the intensity bins, although it tends towards a shape factor of $\xi\approx 0$ at the highest intensities, indicating either a random orientation or transitioning towards perpendicular alignment. This can also be seen qualitatively in Figure~\ref{fig:lic} and the upper panel of Figure~\ref{fig:hro}, which show an approximately north-south-aligned magnetic field, which matches the major axis of the cloud delineated by the FUV contour. The mean \textit{Planck} magnetic field orientation is 163.5$^{\circ}$ E of N, while the Eos cloud has a major axis of $\approx165^{o}$ E of N.

Considering the column density values of the Eos cloud, $<10^{21}$\,cm$^{-2}$ \citep{burkhart2025}, this consistently preferentially parallel orientation is as expected. \citet{Planck2016_xxxv} found a typical transition column density value of $\gtrsim10^{21.7}$\,cm$^{-2}$, above which the magnetic field becomes preferentially perpendicular to the cloud structure. The observed parallel relative orientation at these column densities could indicate the turbulence in Eos is either Alfv\'enic or sub-Alfv\'enic \citep{Planck2016_xxxv}, and that the magnetic field plays a dynamically important role in the evolution of the cloud.

\subsection{Magnetic Field Strength}
\label{sec:results_dcf}

The plane-of-sky magnetic field strength in molecular clouds can be estimated from dust polarization observations using the Davis-Chandrasekhar-Fermi (DCF) method \citep{1951PhRv...81..890D,1953ApJ...118..116C}. We perform this analysis for two regimes in the Eos cloud: across the cloud as a whole, and within the denser CO clump MBM 40. A parametrized version of the DCF method, assuming fully molecular gas is given by
\begin{equation}
  B_{\rm pos}^{\rm DCF}(\mu {\rm G}) \approx 18.6\,Q\sqrt{n(\rm H_{2})({\rm cm}^{-3})}\frac{\Delta v_{\rm NT}(\rm km\, s^{-1})}{\sigma_{\theta}(\rm degree)} \,,
  \label{eq:badf}
\end{equation}
\noindent
 following \citet{crutcher04}, where $n(\rm H_{2}$) is the volume density of molecular hydrogen and $\Delta v_{\rm NT}$ is the FWHM of the non-thermal gas velocity dispersion, given by $\Delta v_{\rm NT}$ = $\sigma_{\rm NT}\sqrt{8\rm{ln}2}$. $\sigma_{\theta}$ is the dispersion of the position angles of the magnetic field vectors. To obtain a dispersion of position angles, we use the structure function approach \citep{2009Hildebrand} over both the entire cloud and the CO core. $Q$ is a correction factor that accounts for variation of the magnetic field along the line of sight and on scales smaller than the beam and is typically taken to be 0.5 \citep[see][]{2001ApJ...546..980O,crutcher04}. 

\begin{figure}
  \includegraphics[width=\columnwidth]{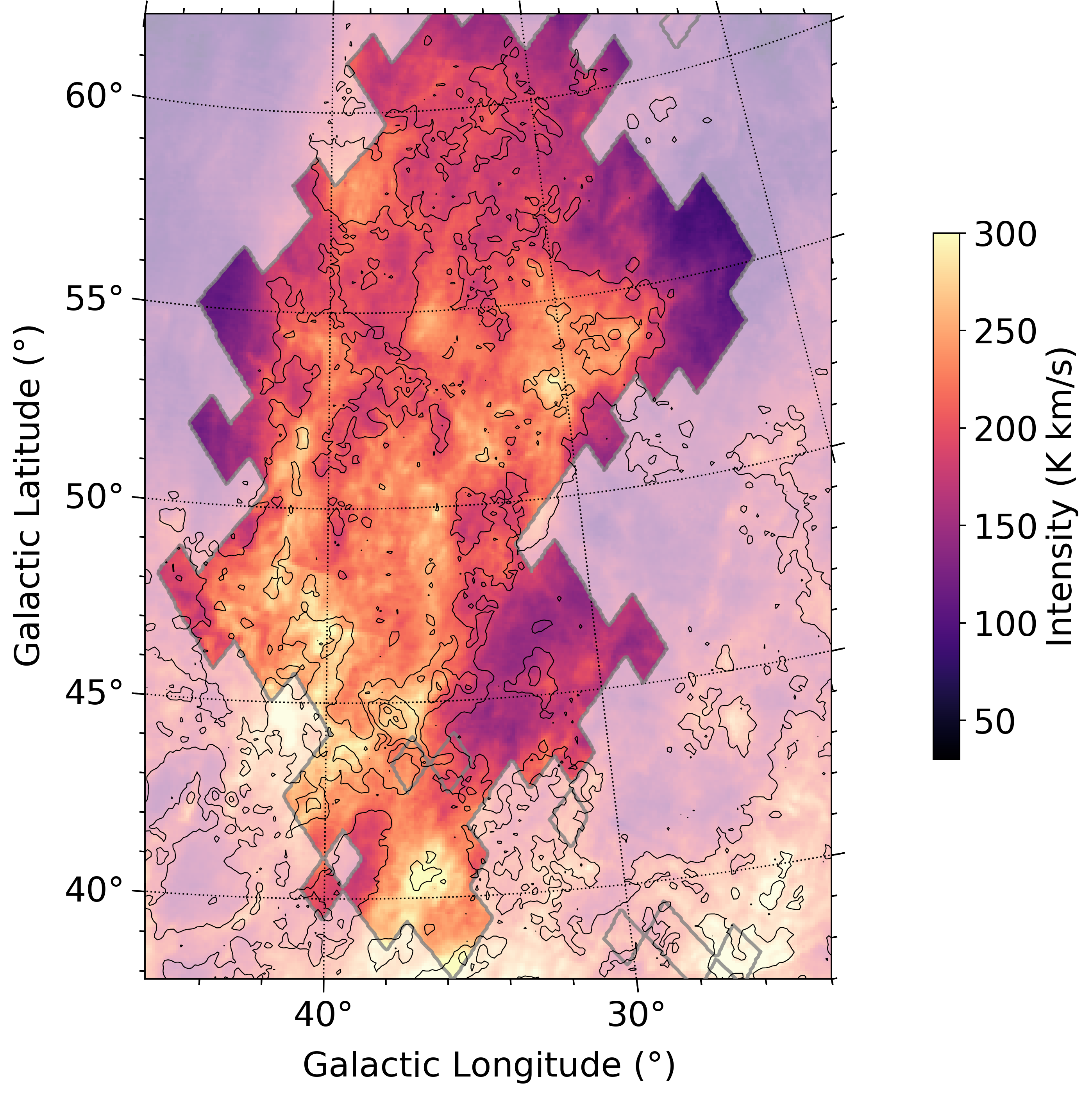}
  \caption{Moment 0 map from the GALFA-H\,{\sc i} data. The grey contour shows the boundary of the cloud from the UV observations. Black contours are the same as Figure~\ref{fig:hro}.}
  \label{fig:HI} 
\end{figure}

\citet{2021Skalidis} provide an alternative formulation of the DCF method
\begin{equation}
  B_{\rm pos}^{\rm ST} = \sqrt{2\pi\rho}\frac{\delta v}{\sqrt{\sigma_{\theta}}} \,,
  \label{eq:stdcf}
\end{equation}
\noindent
which when compared to the classical DCF (see their Eq. 31), it gives that B$_{\rm pos}^{\rm DCF}$ = $Q\sqrt{2/\sigma_{\theta}}$ B$_{\rm pos}^{\rm ST}$. For $Q=0.5$, the classical DCF (B$_{\rm pos}^{\rm DCF}$) is a factor $\approx$0.707$(\sigma_{\theta})^{-0.5}$ greater than the \citet{2021Skalidis} formulation (B$_{\rm pos}^{\rm ST}$) where $\sigma_{\theta}$ is in radians.

\subsubsection{Column and volume densities}
\label{subsubsec:dens}

We use the H$_2$ mass, $M=3.4\times10^{3}\,$M$_{\odot}$, and $r=25.5$\,pc \citep{burkhart2025} to calculate the H$_2$ column and volume density of the entire Eos cloud. The calculated volume and column densities are given in Table~\ref{tab:tab1}. For MBM 40, we use values from the Planck Galactic Cold Clump \citep[PGCC;][]{planck_pgcc} Survey, which identified a dense clump, PGCC G37.52+44.57, within MBM 40. It should be noted that this dense clump is of arcminute-scale, while the MBM 40 cloud identified from CO emission, over which we have analyzed the magnetic field, is of degree-scale. Thus this value for n(H$_2$) is an upper limit. The temperature, column density and volume density values adopted for Eos as a whole and for MBM 40 are all given in Table~\ref{tab:tab1}.

\subsubsection{Velocity dispersion}

 We determine velocity dispersions for the Eos cloud using the H\,{\sc i} measurements of \citet{verschuur1994}, who identified two main H\,{\sc i} components in high Galactic latitude clouds: velocity dispersions in the range of 8--13\,km\,s$^{-1}$, corresponding to material at temperatures of $\sim$2000--3500\,K, and velocity dispersions of $\sim$3\,km\,s$^{-1}$ corresponding to material at temperatures $\sim$200--400\,K. This is similar to the H\,{\sc i} decomposition performed by \citet{soler2018} in the Orion-Eridanus superbubble, with velocity dispersions of $\sim$2\,km\,s$^{-1}$ identified to be corresponding to the CNM and dispersions $>$8\,km\,s$^{-1}$ corresponding to the warm neutral medium (WNM).

In H\,\textsc{i} emission in the vicinity of MBM 40, \citet{verschuur1994} show two velocity components, one at $\sim$3\,km\,s$^{-1}$ and another at $\sim$8\,km\,s$^{-1}$, with the $\sim$3\,km\,s$^{-1}$ component being significantly the stronger of the two (i.e. more lines show dispersions of $\sim$3\,km\,s$^{-1}$).

If we consider the CNM component, the total velocity dispersion from \citet{verschuur1994} is given as $\sim$3\,km/s for material at 300--400\,K. For an order-of-magnitude estimation, we consider the temperature of the CNM material in MBM 40 to be 350$\pm$50\,K based on the \citet{verschuur1994} decomposition, and the velocity dispersion to be 3$\pm$1\,km\,s$^{-1}$. Using $T=350$\,K, we can calculate the thermal component to be 1.7\,km\,s$^{-1}$. Removing the thermal component through quadrature subtraction gives a non-thermal velocity dispersion of 2.5\,km\,s$^{-1}$, equivalent to a FWHM linewidth of 5.8\,km\,s$^{-1}$ for the CNM material. 

Similarly, if we consider the WNM component, the velocity dispersion from \citet{verschuur1994} is given as $\sim$8\,km\,s$^{-1}$ for material at 2000\,K. The thermal component is 4.1\,km\,s$^{-1}$ and removing this gives a non-thermal velocity dispersion of 6.9\,km/s, or a non-thermal FWHM linewidth of 16.2\,km/s for the WNM material. 

Inspection of the GALFA-H\,{\sc i} data cubes for the Eos molecular cloud indeed show these multiple-component Gaussians are present across the entire cloud, and are not restricted to the vicinity of MBM40. We therefore choose to extend the calculated non-thermal linewidths above from MBM40 out to the entire Eos cloud. 

In general, there is a strong correlation found between magnetic field and CNM structures, based on dust polarized emission and H\,{\sc i} data, respectively \citep{clark2014,Planck2016_xxxii}. This is in contrast to a weak correlation with WNM material. In Eos, the H\,{\sc i} emission traces the dust emission quite well (see Figure~\ref{fig:HI}) and so we elect to use the CNM linewidths when calculating magnetic field strength. We note that larger linewidths arising from the WNM would produce higher magnetic field strengths (see equation~\ref{eq:badf}), although this might be mitigated by WNM material being less dense than CNM material. 

For MBM 40 itself, due to its higher density and presence of CO, we determine the velocity linewidth from CO data. \citet{2022ApJS..262....5D} include MBM 40 in their CO survey of the northern hemisphere and find a linewidth of 0.64\,km\,s$^{-1}$ (MBM 40 corresponds to their Cloud 32). This is in good agreement with \citet{shore03} who find $^{12}$CO FWHM line widths of 0.8--0.9\,km\,s$^{-1}$ and $^{13}$CO FWHM line widths of 0.4--0.6\,km\,s$^{-1}$. Assuming a temperature of 15.7\,K \citep{planck_pgcc}, the thermal line width is negligible ($\approx$0.01\,km\,s$^{-1}$).

\subsubsection{Magnetic field dispersion}

To determine the dispersion in polarization position angle, $\sigma_{\theta}$, in the presence of magnetic fields which may show ordered variation, we {\rm use} the structure function approach \citep[SF;][]{2009Hildebrand}. For each pair of magnetic field vectors, their distance and angle difference is calculated; a histogram of average angle difference is then plotted and fitted with a quadratic function to measure underlying angle dispersion.  While measurement uncertainty on position angle is typically subtracted in such analyses, \textit{Planck} Commander Stokes $Q$ and $U$ maps are not provided with uncertainties. While this means that the position angle dispersions which we calculate are formally upper limits, we expect the effect on our measured magnetic field strengths to be small.

\begin{figure}[!t]
  \centering
  \includegraphics[width=\columnwidth]{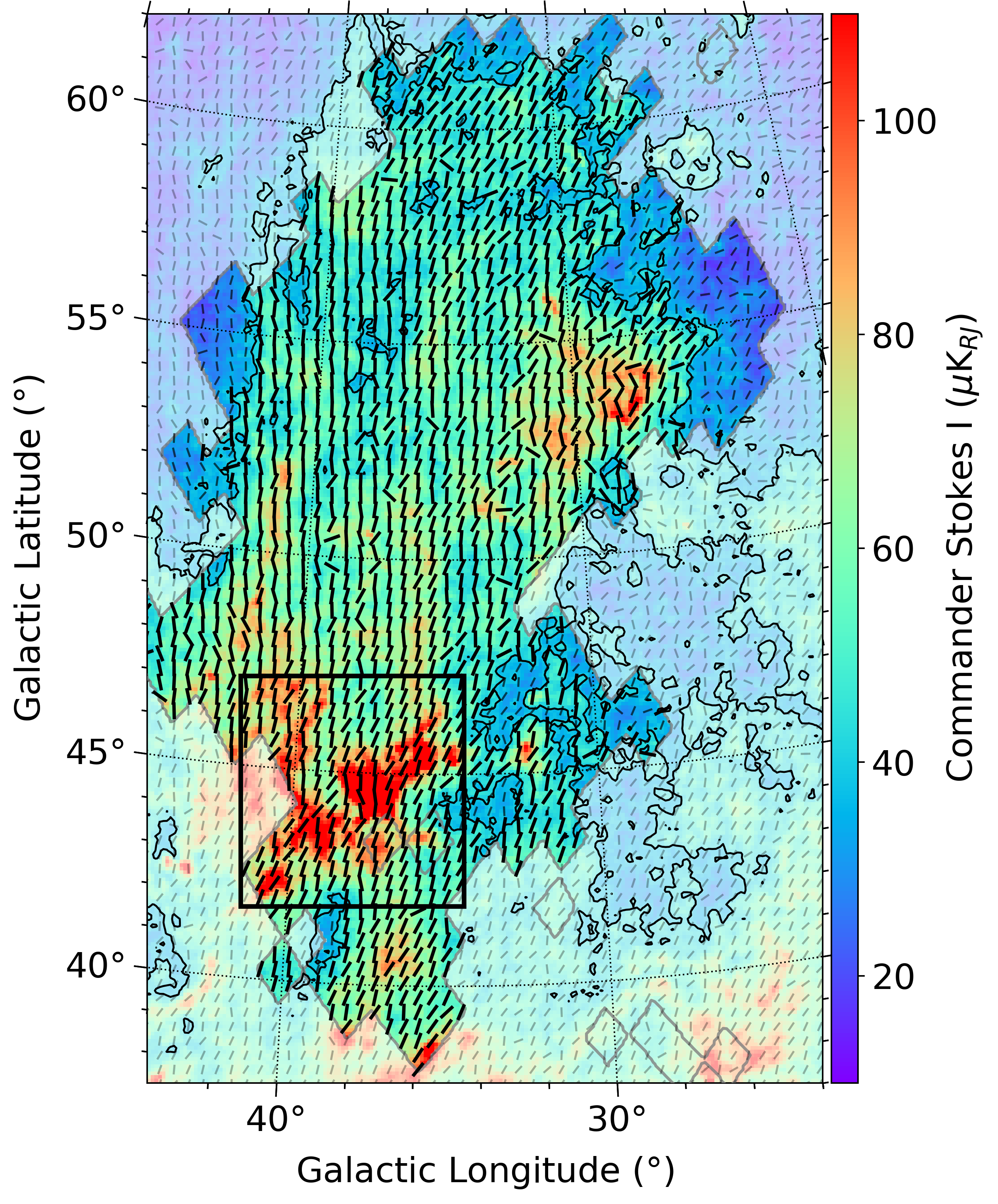} \\
  \includegraphics[width=0.9\columnwidth]{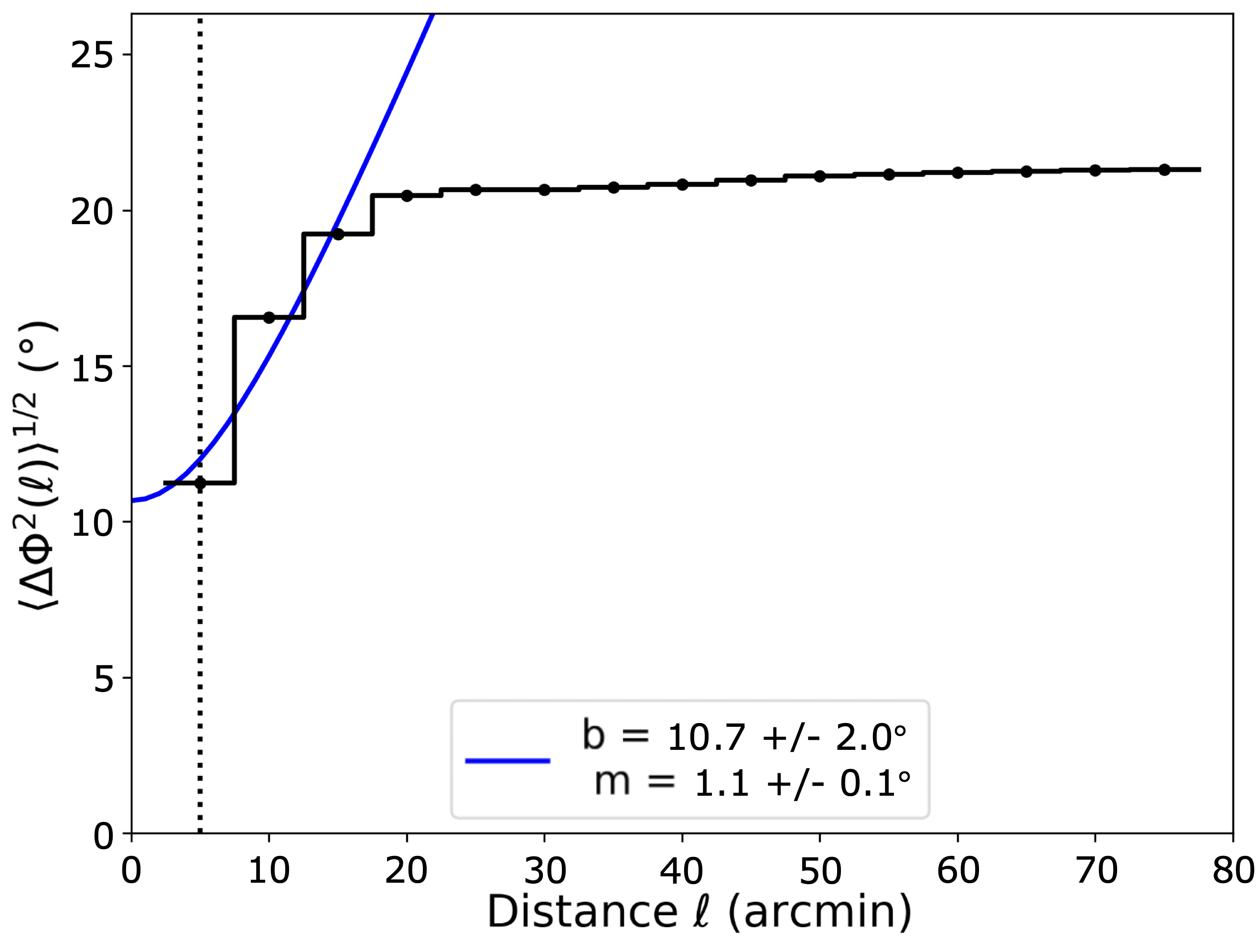}
  \caption{Background of the left panel is the \textit{Planck} Stokes \textit{I} map. Selected vectors are in bold and black, with a selection criteria of \textit{I}$>$40$\mu$K$_{\rm RJ}$. Non-selected vectors are fainter. The region shown in Figure~\ref{fig:sf2} is shown in the black box. The structure function of the selected vectors is shown on the right \citep{2009Hildebrand}.}
  \label{fig:sf1} 
\end{figure}

\begin{figure}[!t]
  \centering
  \includegraphics[width=\columnwidth]{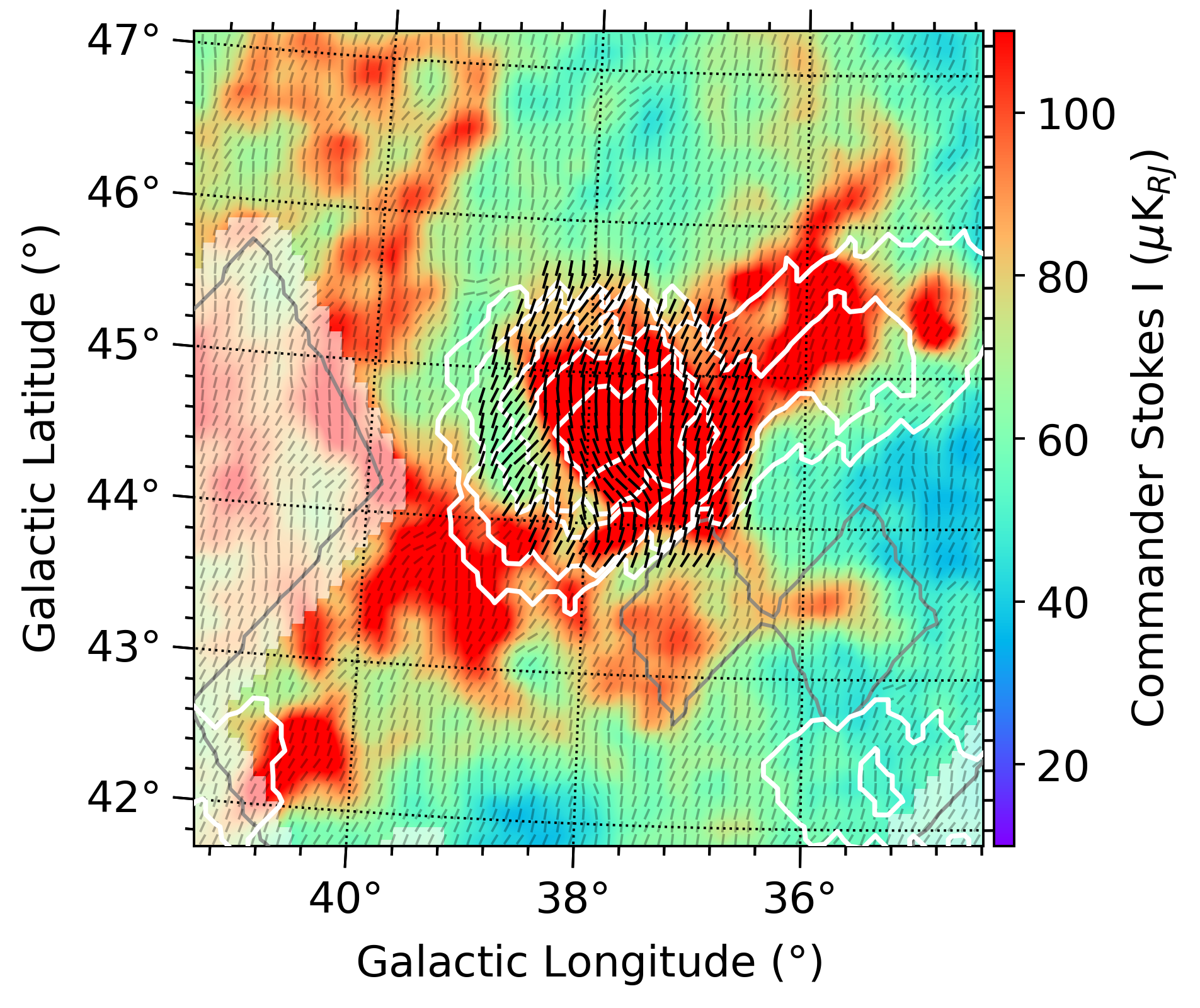}
  \includegraphics[width=0.9\columnwidth]{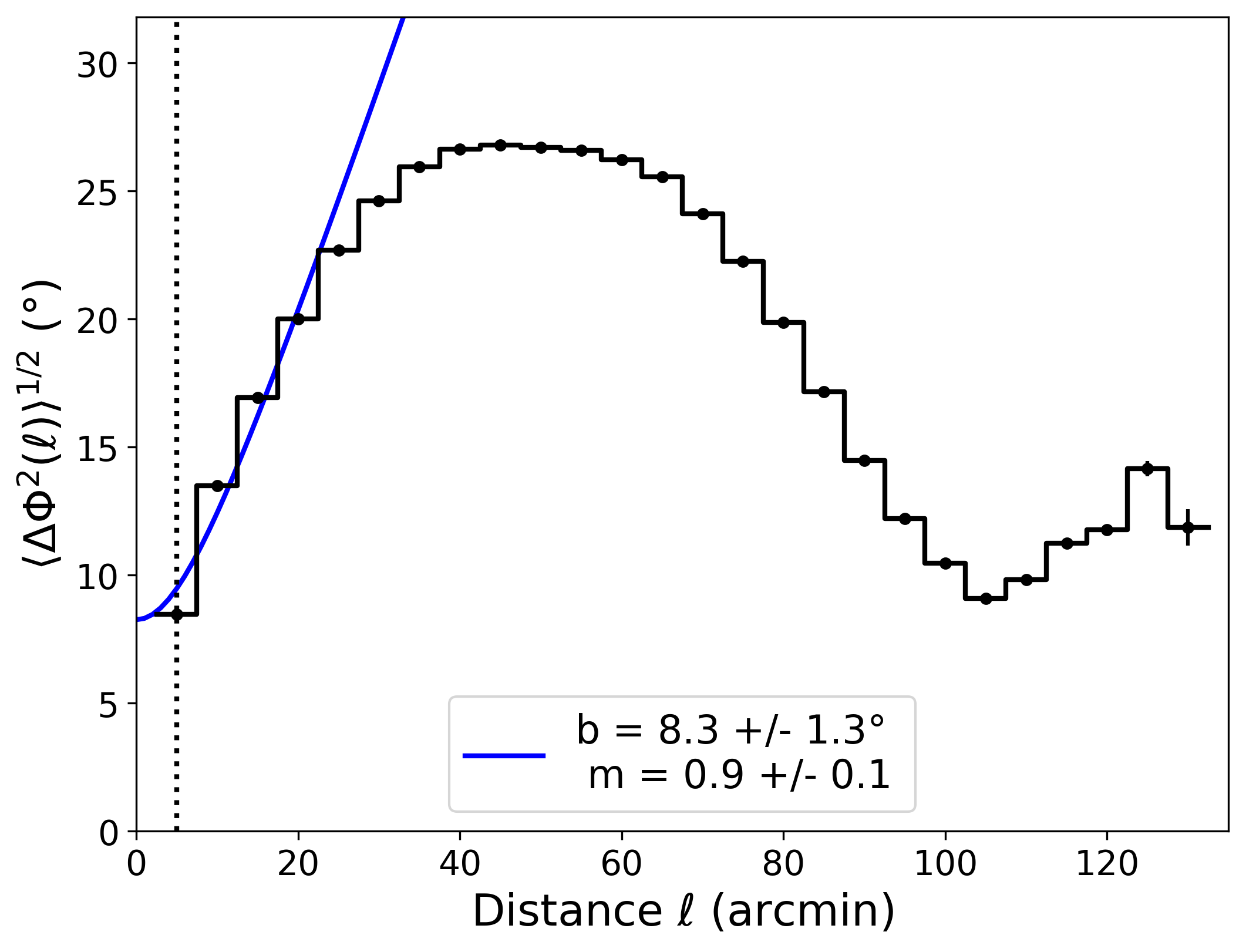}
  \caption{Same as Figure~\ref{fig:sf1}. Planck CO (1-0) contours are overlaid in white. Selected vectors are plotted (in black) within the masked area from the CO emission found in \citet{2022ApJS..262....5D}. The structure function of the black vectors is shown on the right \citep{2009Hildebrand}.}
  \label{fig:sf2} 
\end{figure}

 The two contributions to the mean angle differences that we measure are therefore turbulent dispersion, which we wish to characterize, and large-scale field variation. We fit the angular dispersion function for the whole Eos cloud (see Figure~\ref{fig:sf1}) and for MBM 40 (see Figure~\ref{fig:sf2}) with the quadratic function
\begin{equation}
  \langle \Delta\Phi^{2}(\ell)\rangle = b^{2} + m^2\ell^2,
\end{equation}
 where $b^{2}$ is the contribution due to the turbulent dispersion of the magnetic field, and $m^{2}\ell^{2}$ characterizes the large-scale field variation as a function of distance $\ell$. The angular dispersion is then given by
\begin{equation}
\sigma_{\theta} = \frac{b}{\sqrt{2-b^{2}}},
\end{equation}
where all values are given in radians. The results are shown in Figures~\ref{fig:sf1} and \ref{fig:sf2} and given in Table~\ref{tab:tab1}. We note the derived angular dispersion is approximately equal to the mean value of S, the polarization angle dispersion function (see Figure~\ref{fig:pvsi}).

\subsubsection{Magnetic field strengths in the Eos cloud}

Using the values listed in Table~\ref{tab:tab1}, we calculate a classical DCF magnetic field strength (B$^{DCF}_{\rm pos}$) of $6.0\pm3.4$\,$\mu$G for the CNM component of the Eos molecular cloud. For the MBM 40 molecular cloud, we calculate a slightly higher magnetic field strength of $12.0\pm4.3$\,$\mu$G, although this value agrees within uncertainty with our measurement for the Eos cloud as a whole. 

For $\sigma_{\theta}$ of 7.6$^\circ$ and 5.9$^\circ$, B$^{ST}_{\rm pos}$ is a factor 1.94 and 2.02 less than B$^{DCF}_{\rm pos}$ respectively.

\begin{table}
  \caption{Values used in, and results of, our magnetic field strength analysis for the Eos cloud as a whole and for the denser MBM 40 region. Uncertainties on each quantity are given in brackets. B$_{\rm pos}^{DCF}$ refers to Eq.~\ref{eq:badf} and B$_{\rm pos}^{ST}$ refers to Eq.~\ref{eq:stdcf}}
  \begin{tabular}{ccc} \hline
 & Eos CNM & MBM 40 \\ \hline
T (K) & 350 (50) & 15.7 (2.2)$^{\rm c}$ \\
N(H$_2$) (10$^{20}$ cm$^{-2}$) & 0.75 (0.30) & 1.5 (0.6)$^{\rm c}$ \\
n(H$_2$) (cm$^{-3}$) & 0.71 (0.28) & 140 (63)$^{\rm c}$ \\
H\,{\sc i} $\sigma$v$^{\rm a}$ (km/s) & 3 (1) & -- \\
H\,{\sc i} $\sigma$v$_{\rm T}$ (km/s) & 1.7 (0.5) & -- \\
H\,{\sc i} $\Delta$v$_{\rm NT}$ (km/s) & 5.8 (3.0) & -- \\
CO $\Delta$v$_{\rm NT}$$^{\rm b}$ (km/s) & -- & 0.64 (0.15) \\
b ($^\circ$) & 10.7 (2) & 8.3 (1.3) \\
$\sigma_{\theta}$ ($^\circ$) & 7.6 (1.4) & 5.9 (0.9) \\
B$_{\rm pos}^{DCF}$ ($\mu$G) & 6.0 (3.4) & 12.0 (4.3) \\
B$_{\rm pos}^{ST}$ ($\mu$G) &  3.1 (1.8) &  5.9 (2.2) \\
$\lambda^{DCF}$ & 0.10 (0.07) & 0.10 (0.05) \\
$\mathcal{M}_{\rm A}^{DCF}$ & 0.38 (0.17) & 0.29 (0.15) \\ \hline
  \end{tabular}
  \begin{center}
  a. Values from \citet{verschuur1994} \\
  b. From \citet{2022ApJS..262....5D} \\ 
  c. From \citet{planck_pgcc}.
  \end{center}
  \label{tab:tab1}
\end{table}

\section{Discussion}
\label{sec:dis}

\subsection{Comparison with the Local Bubble}

The Local Bubble (LB) is thought to have originated from several supernovae by stars in the same cluster that the Sun was formed in \citep{smith2001,zucker22}. While vented to high Galactic latitudes, the state of the diffuse gas in the LB has been a subject of ongoing debate. Diffuse soft X-ray emission \citep{bowyer1968,mccammon1983,snowden1998} points to an, at least partial, hot interior (T$\sim$ 10$^{6}$ K). If collisional ionization equilibrium is assumed, a thermal pressure of P/$k$=2nT$\approx$15,000\,K\,cm$^{-3}$ is implied \citep{snowden1998}, where the factor 2 occurs as the plasma is fully ionized. Extreme Ultraviolet Explorer (EUVE) observations \citep{bowyer1995} yield a slightly higher value of P/$k\sim19,000$\,K\,cm$^{-3}$.

If the excitation of the gas is out of equilibrium (not yet equilibrated from the last SNR passage) the gas may be at lower thermal pressure \citep{br1994}. \citet{jenkins2002} used observations of the UV fine-structure lines of C\,{\sc i} to derive thermal pressures for the gas towards three stars in the LB of P/$k<10,000$\,K\,cm$^{-3}$, and notes that the discrepancy between the X-ray/EUVE and C\,{\sc i} pressures may be explained by significant magnetic pressure contributions. \citet{andersson06} used optical polarization measurements of the Local Bubble wall (and the classical form of the DCF relation), in the direction of the Southern Coalsack, to derive a magnetic field of B=$8^{+5}_{-3}$\,$\mu$G, equivalent to a magnetic pressure of pressure of P$_{\rm B}$/k$\sim18,000$\,K\,cm$^{-3}$.

The magnetic field strength that we derive in Eos is also $\approx$6$\pm$3\,$\mu$G. Finding this magnetic field in a separate area of the LB suggests that the LB has a magnetic field throughout which is strong enough to provide the pressure support needed. 

\subsection{A Globally Important Magnetic Field?}
 
In addition to significant magnetic pressure contributions in the LB, the magnetic field may play a role in governing dynamics of the Eos molecular cloud.

Section~\ref{subsec:hro} showed that the magnetic field is predominantly parallel to the structure of Eos. This occurs in magnetically-dominated flux-frozen and non-self-gravitating regimes where gas is expected to flow along magnetic field lines, rather than the gas dragging the field lines as it moves. This also suggests that the turbulence in Eos is either Alfv\'enic or sub-Alfv\'enic \citep{Planck2016_xxxv}.

We calculate the Alfv\'enic Mach number as
\begin{equation}
  \mathcal{M}_{\rm A} = 1.74\times10^{-2} \sqrt{2} \; \frac{\sigma_{\theta}(\rm degree)}{Q} \;,
  \label{eq:machsimp}
\end{equation}
\noindent
continuing to take $Q=0.5$. We have also included a factor of $\sqrt{2}$ in Equation~\ref{eq:machsimp}, as recommended by \citet{Heiles_2005} to account for the velocity dispersion being a one-dimensional measurement along the LOS, while the angle dispersion is measured in the POS. We also note that this value should consider the 3D dispersion but DCF only allows determination of the POS component. Since $\mathcal{M}_{\rm A}\propto$ B$_{tot}^{-1}$ and B$_{\rm pos}$=B$_{\rm tot}$\,$sin\theta$, our 2D $\mathcal{M}_{\rm A}$ can be considered an upper limit, although, unless the magnetic field has a very substantial LOS component, the value calculated from B$_{\rm pos}$ will only differ by a factor of a few. We find Alfv\'enic Mach numbers for Eos and MBM 40 of 0.38$\pm$0.17 and 0.3$\pm$0.1 respectively. In both cases $\mathcal{M}_{\rm A}<1$ which suggests the clouds are sub-Alfv\'enic, in line with the results of the HRO analysis.
 
We can also compare the dynamic importance of magnetic fields and gravity in Eos, which is typically parametrized using the mass-to-flux ratio, $\lambda$, which is given by
\begin{equation}
\lambda= \frac{(M/\Phi)_{\rm obs}}{(M/\Phi)_{\rm crit}} = 7.6\times10^{-21}\frac{N_{\rm H_{2}}(\rm cm^{-2})}{B_{\rm pos}(\mu \rm G)} \, \, ,
\label{eq:lam}
\end{equation}
\noindent
\citep{crutcher04}. If we use the average column density of the cloud and the calculated magnetic field strength of $6.0\pm3.4$\,$\mu$G, we find $\lambda=0.10\pm0.07$. Similar to the above argument concerning $\mathcal{M}_{\rm A}$, $\lambda\propto$ B$_{tot}^{-1}$ and so our value here, calculated using the 2D B$_{pos}$, can be considered an upper limit.\footnote{Note that \citet{crutcher04} derive a statistical correction factor $\lambda=\lambda_{\rm obs}/3$, but since this is applicable only to an ensemble of measurements, we do not apply it here.} A value of $\lambda<1$ is referred to as magnetically subcritical, indicating that the magnetic field can support the cloud against gravitational collapse. The CNM component of the Eos molecular cloud is therefore magnetically subcritical. We perform the same calculation for the MBM 40 molecular cloud and find $\lambda=0.10\pm0.05$. In both cases, $\lambda$ remains sub-critical if B$_{\rm pos}^{DCF}$ is replaced with B$_{\rm pos}^{ST}$.

There are many indicators that the Eos cloud is not gravitationally unstable: \citet{burkhart2025} find that the cloud may be actively evaporating, while \citet{saxena2025} find no indication of any past or present star formation in the cloud. Our subcritical mass-to-flux ratios further support this picture, and suggest that even in the absence of evaporative effects, the cloud would still be supported against collapse by its magnetic field.

The size and volume density of the Eos cloud and the MBM 40 sub-cloud vary by approximately one and two orders of magnitude respectively. However, the magnetic field, and the mass-to-flux ratio, remain consistent between them. This is as predicted by \citet{crutcher2010}, in whose model at low H\,{\sc i} volume densities ($n$(H\,{\sc i})$\lesssim300$\,cm$^{-3}$), the maximum magnetic field strength remains constant as $n$ increases. The critical volume density in terms of molecular hydrogen, H$_2$ is $n(\rm H_{2})\lesssim150$\,cm$^{-3}$, which is true of both the Eos cloud and MBM 40, especially when considering the n(H$_2$) of MBM 40 is an upper limit (see Section~\ref{subsubsec:dens}). The constant magnetic field strength in this model is $\approx$10\,$\mu$G \citep{crutcher2010}, which is within the error bounds of our measured magnetic field strengths (see Table~\ref{tab:tab1}). This regime of the $B$ versus $n$ relation is where the magnetic field is considered to be dynamically important \citep[][and references therein]{crutcher2010}, before the cloud becomes self-gravitating.

The \citet{crutcher2010} model applies to total magnetic field strengths, while we report POS field strengths here. To account for this, we estimate that the magnetic field is parallel to the cloud structure of Eos (a likely valid assumption, see Section~\ref{subsec:hro}), then an estimation of the cloud inclination will provide an estimate of the magnetic field inclination. We take the angular size of the cloud to be $\sim$20$^{\circ}$ which is equivalent to a POS extent of $\sim$42\,pc. The depth of the cloud we take to also be 42\,pc \citep[an upper limit, see Figure 3 of][]{burkhart2025}, which gives an inclination angle of 45$^{\circ}$. Then B$_{\rm pos}$=B$_{\rm tot}$\,$sin\theta$ so B$_{\rm tot}$=$\sqrt{2}$B$_{\rm pos}\approx1.4$B$_{\rm pos}$. If we assume a smaller depth, e.g. 30\,pc \citep[not including the first or final layer of the Eos cloud, again see Figure 3 of][]{burkhart2025}, then the inclination angle is 54$^{\circ}$ and B$_{\rm tot}\approx1.2$B$_{\rm pos}$. This shows lower depths mean a higher inclination angle which tends towards B$_{\rm tot}\approx$B$_{\rm pos}$. The effects of these correction factors are smaller than our uncertainties, and so accounting for the LOS component of the magnetic field does not change our results.

These arguments show that the magnetic field in the Eos and MBM40 clouds are dynamically important for every metric we consider: they are aligned with the major axis of the cloud, subcritical, sub-Alfv\'enic, and do not vary as a function of gas density. Our results are therefore consistent both with the (absence of) star formation history in the Eos cloud \citep{saxena2025}, and also with the paradigm in which molecular clouds form from a magnetically subcritical CNM in which magnetic fields are dynamically important \citep{crutcher2010,Planck2016_xxxv}. With the current destruction (evaporation) of Eos \citep{burkhart2025}, the magnetic field may no longer be playing as significant a role in the evolution of the cloud as they may have done in its formation. If the Eos cloud is evaporating due to interactions with hot gas, the magnetic field could play a role in slowing down the evaporation by restricting the transport of hot electrons from the coronal gas to the CNM, or even the WNM \citep[e.g.][]{andersson04,fox06}.

\section{Conclusions}
\label{sec:con}

We have presented observations of the polarization from the  newly discovered Eos molecular cloud using \textit{Planck} dust emission and starlight extinction measurements. Based on these, we find a well-ordered, linear magnetic field aligned with the north-south major axis of the cloud, which is dynamically important according to every metric that we consider and consistent with the Eos cloud having formed from a strongly magnetized CNM.

We investigated the dust polarization properties of the molecular cloud and find power law indices of $-0.61\pm0.08$ and $-0.53\pm0.03$ for starlight polarization and dust emission polarization, respectively, suggesting constant grain alignment efficiency, in a turbulent medium, in Eos. We find a shallower power law index, 0.82, than predicted by the analytical model of \citet{planckxii} when fitting the polarization angle dispersion function, $\mathcal{S}$, versus polarization fraction, p$_{frac}$, potentially suggesting a less disorganized magnetic field. We also find nearly constant $\mathcal{S} \times p_{frac}$ with increasing intensity suggesting the the observed grain alignment efficiency in this cloud does not vary significantly with local conditions.

We used HRO and DCF analyses to investigate the role, and dynamic importance, of magnetic fields within Eos. Using the DCF method, we derived magnetic field strengths of $6\pm3$\,$\mu$G and $12\pm4$\,$\mu$G for Eos and MBM 40, respectively. We find similar values within errors using other formulations and methods of deriving the magnetic field strength. These values are consistent with the previous estimate of magnetic field strength in the Local Bubble and consistent with a picture of magnetic pressure support within the Local.

We find four strong results supporting the conclusion of a dynamically important magnetic field in the formation of the Eos cloud:
\vspace{6pt}
\begin{enumerate}[noitemsep,nosep,leftmargin=*]
\item The HRO shows that the magnetic field is preferentially parallel to the density structure, which occurs in magnetically-dominated regimes where material is expected to flow along magnetic field lines.
\item The material in Eos is found to be sub-Alfv\'enic, a conclusion which aligns with the preferentially parallel orientation of the magnetic field and density structure.
\item The Eos cloud has a strongly subcritical mass-to-flux ratio, indicating that the magnetic field can support the cloud against gravitational collapse.
\item The magnetic field strength is consistent between the Eos cloud as a whole and the higher-density MBM 40 region, as predicted by the \citet{crutcher2010} model for magnetic fields at relatively low densities in molecular clouds that have formed from a magnetically subcritical CNM.
\end{enumerate}
\vspace{6pt}
 All of these findings are fully consistent with the observed lack of star formation in the Eos cloud, and with our expectations for a low-density, non-self-gravitating, molecular cloud which has formed from a CNM in which magnetic fields are dynamically important.

\begin{acknowledgements}
We thank the anonymous reviewer for their comments, which improved the science and clarity of the manuscript.
J.K. is currently supported by the Royal Society under grant number RF\textbackslash ERE\textbackslash231132, as part of project URF\textbackslash R1\textbackslash211322.
K.P. is a Royal Society University Research Fellow, supported by grant number URF\textbackslash R1\textbackslash211322.
B.B. acknowledges support from NSF grant AST-2009679 and NASA grant No. 80NSSC20K0500.
B.B. is grateful for the generous support by the David and Lucile Packard Foundation and Alfred P. Sloan Foundation. 
B.B. thanks the Center for Computational Astrophysics (CCA) of the Flatiron Institute and the Mathematics and Physical Sciences (MPS) division of the Simons Foundation for support.
T.E.D. acknowledges support for this work provided by NASA through the NASA Hubble Fellowship Program grant No. HST-HF2-51529 awarded by the Space Telescope Science Institute, which is operated by the Association of Universities for Research in Astronomy, Inc., for NASA, under contract NAS 5-26555.
T.J.H acknowledges UKRI guaranteed funding for a Horizon Europe ERC consolidator grant (EP/Y024710/1) and a Royal Society Dorothy Hodgkin Fellowship.
\end{acknowledgements}

\vspace{5mm}
\facilities{\textit{Planck} Observatory}


\software{Astropy \citep{2013A&A...558A..33A,2018AJ....156..123A}, SciPy \citep{2020SciPy-NMeth}}

\bibliography{eos_b}{}

\begin{thebibliography}{}
\expandafter\ifx\csname natexlab\endcsname\relax\def\natexlab#1{#1}\fi
\providecommand{\url}[1]{\href{#1}{#1}}
\providecommand{\dodoi}[1]{doi:~\href{http://doi.org/#1}{\nolinkurl{#1}}}
\providecommand{\doeprint}[1]{\href{http://ascl.net/#1}{\nolinkurl{http://ascl.net/#1}}}
\providecommand{\doarXiv}[1]{\href{https://arxiv.org/abs/#1}{\nolinkurl{https://arxiv.org/abs/#1}}}

\bibitem[{{Andersson} {et~al.}(2004){Andersson}, {Knauth}, {Snowden}, {Shelton}, \& {Wannier}}]{andersson04}
{Andersson}, B.~G., {Knauth}, D.~C., {Snowden}, S.~L., {Shelton}, R.~L., \& {Wannier}, P.~G. 2004, \apj, 606, 341, \dodoi{10.1086/382861}

\bibitem[{{Andersson} {et~al.}(2015){Andersson}, {Lazarian}, \& {Vaillancourt}}]{2015Andersson}
{Andersson}, B.~G., {Lazarian}, A., \& {Vaillancourt}, J.~E. 2015, \araa, 53, 501, \dodoi{10.1146/annurev-astro-082214-122414}

\bibitem[{{Andersson} \& {Potter}(2006)}]{andersson06}
{Andersson}, B.~G., \& {Potter}, S.~B. 2006, \apjl, 640, L51, \dodoi{10.1086/503199}

\bibitem[{{Astropy Collaboration} {et~al.}(2013){Astropy Collaboration}, {Robitaille}, {Tollerud}, {Greenfield}, {Droettboom}, {Bray}, {Aldcroft}, {Davis}, {Ginsburg}, {Price-Whelan}, {Kerzendorf}, {Conley}, {Crighton}, {Barbary}, {Muna}, {Ferguson}, {Grollier}, {Parikh}, {Nair}, {Unther}, {Deil}, {Woillez}, {Conseil}, {Kramer}, {Turner}, {Singer}, {Fox}, {Weaver}, {Zabalza}, {Edwards}, {Azalee Bostroem}, {Burke}, {Casey}, {Crawford}, {Dencheva}, {Ely}, {Jenness}, {Labrie}, {Lim}, {Pierfederici}, {Pontzen}, {Ptak}, {Refsdal}, {Servillat}, \& {Streicher}}]{2013A&A...558A..33A}
{Astropy Collaboration}, {Robitaille}, T.~P., {Tollerud}, E.~J., {et~al.} 2013, \aap, 558, A33, \dodoi{10.1051/0004-6361/201322068}

\bibitem[{{Astropy Collaboration} {et~al.}(2018){Astropy Collaboration}, {Price-Whelan}, {Sip{\H{o}}cz}, {G{\"u}nther}, {Lim}, {Crawford}, {Conseil}, {Shupe}, {Craig}, {Dencheva}, {Ginsburg}, {VanderPlas}, {Bradley}, {P{\'e}rez-Su{\'a}rez}, {de Val-Borro}, {Aldcroft}, {Cruz}, {Robitaille}, {Tollerud}, {Ardelean}, {Babej}, {Bach}, {Bachetti}, {Bakanov}, {Bamford}, {Barentsen}, {Barmby}, {Baumbach}, {Berry}, {Biscani}, {Boquien}, {Bostroem}, {Bouma}, {Brammer}, {Bray}, {Breytenbach}, {Buddelmeijer}, {Burke}, {Calderone}, {Cano Rodr{\'\i}guez}, {Cara}, {Cardoso}, {Cheedella}, {Copin}, {Corrales}, {Crichton}, {D'Avella}, {Deil}, {Depagne}, {Dietrich}, {Donath}, {Droettboom}, {Earl}, {Erben}, {Fabbro}, {Ferreira}, {Finethy}, {Fox}, {Garrison}, {Gibbons}, {Goldstein}, {Gommers}, {Greco}, {Greenfield}, {Groener}, {Grollier}, {Hagen}, {Hirst}, {Homeier}, {Horton}, {Hosseinzadeh}, {Hu}, {Hunkeler}, {Ivezi{\'c}}, {Jain}, {Jenness}, {Kanarek}, {Kendrew}, {Kern}, {Kerzendorf}, {Khvalko}, {King}, {Kirkby}, {Kulkarni},
  {Kumar}, {Lee}, {Lenz}, {Littlefair}, {Ma}, {Macleod}, {Mastropietro}, {McCully}, {Montagnac}, {Morris}, {Mueller}, {Mumford}, {Muna}, {Murphy}, {Nelson}, {Nguyen}, {Ninan}, {N{\"o}the}, {Ogaz}, {Oh}, {Parejko}, {Parley}, {Pascual}, {Patil}, {Patil}, {Plunkett}, {Prochaska}, {Rastogi}, {Reddy Janga}, {Sabater}, {Sakurikar}, {Seifert}, {Sherbert}, {Sherwood-Taylor}, {Shih}, {Sick}, {Silbiger}, {Singanamalla}, {Singer}, {Sladen}, {Sooley}, {Sornarajah}, {Streicher}, {Teuben}, {Thomas}, {Tremblay}, {Turner}, {Terr{\'o}n}, {van Kerkwijk}, {de la Vega}, {Watkins}, {Weaver}, {Whitmore}, {Woillez}, {Zabalza}, \& {Astropy Contributors}}]{2018AJ....156..123A}
{Astropy Collaboration}, {Price-Whelan}, A.~M., {Sip{\H{o}}cz}, B.~M., {et~al.} 2018, \aj, 156, 123, \dodoi{10.3847/1538-3881/aabc4f}

\bibitem[{{Barreto-Mota} {et~al.}(2021){Barreto-Mota}, {de Gouveia Dal Pino}, {Burkhart}, {Melioli}, {Santos-Lima}, \& {Kadowaki}}]{2021MNRAS.503.5425B}
{Barreto-Mota}, L., {de Gouveia Dal Pino}, E.~M., {Burkhart}, B., {et~al.} 2021, \mnras, 503, 5425, \dodoi{10.1093/mnras/stab798}

\bibitem[{{Berdyugin} {et~al.}(2014){Berdyugin}, {Piirola}, \& {Teerikorpi}}]{berdyugin2014}
{Berdyugin}, A., {Piirola}, V., \& {Teerikorpi}, P. 2014, \aap, 561, A24, \dodoi{10.1051/0004-6361/201322604}

\bibitem[{{Bowyer} {et~al.}(1968){Bowyer}, {Field}, \& {Mack}}]{bowyer1968}
{Bowyer}, C.~S., {Field}, G.~B., \& {Mack}, J.~E. 1968, \nat, 217, 32, \dodoi{10.1038/217032a0}

\bibitem[{{Bowyer} {et~al.}(1995){Bowyer}, {Lieu}, {Sidher}, {Lampton}, \& {Knude}}]{bowyer1995}
{Bowyer}, S., {Lieu}, R., {Sidher}, S.~D., {Lampton}, M., \& {Knude}, J. 1995, \nat, 375, 212, \dodoi{10.1038/375212a0}

\bibitem[{{Breitschwerdt} \& {Schmutzler}(1994)}]{br1994}
{Breitschwerdt}, D., \& {Schmutzler}, T. 1994, \nat, 371, 774, \dodoi{10.1038/371774a0}

\bibitem[{{Burkhart} {et~al.}(2015){Burkhart}, {Lee}, {Murray}, \& {Stanimirovi{\'c}}}]{2015ApJ...811L..28B}
{Burkhart}, B., {Lee}, M.-Y., {Murray}, C.~E., \& {Stanimirovi{\'c}}, S. 2015, \apjl, 811, L28, \dodoi{10.1088/2041-8205/811/2/L28}

\bibitem[{{Burkhart} {et~al.}(2025){Burkhart}, {Dharmawardena}, {Bialy}, {Haworth}, {Cruz Aguirre}, {Jo}, {Andersson}, {Chung}, {Edelstein}, {Grenier}, {Hamden}, {Han}, {Hoadley}, {Lee}, {Min}, {M{\"u}ller}, {Pattle}, {Peek}, {Pleiss}, {Schiminovich}, {Seon}, {Wilson}, \& {Zucker}}]{burkhart2025}
{Burkhart}, B., {Dharmawardena}, T.~E., {Bialy}, S., {et~al.} 2025, arXiv e-prints, arXiv:2504.17843.
\newblock \doarXiv{2504.17843}

\bibitem[{Cabral \& Leedom(2023)}]{lic_cabral}
Cabral, B., \& Leedom, L.~C. 2023, Imaging Vector Fields Using Line Integral Convolution, 1st edn. (New York, NY, USA: Association for Computing Machinery).
\newblock \url{https://doi.org/10.1145/3596711.3596752}

\bibitem[{{Chandrasekhar} \& {Fermi}(1953)}]{1953ApJ...118..116C}
{Chandrasekhar}, S., \& {Fermi}, E. 1953, \apj, 118, 116, \dodoi{10.1086/145732}

\bibitem[{{Clark} {et~al.}(2014){Clark}, {Peek}, \& {Putman}}]{clark2014}
{Clark}, S.~E., {Peek}, J.~E.~G., \& {Putman}, M.~E. 2014, \apj, 789, 82, \dodoi{10.1088/0004-637X/789/1/82}

\bibitem[{{Clayton} {et~al.}(2003){Clayton}, {Wolff}, {Sofia}, {Gordon}, \& {Misselt}}]{clayton2003}
{Clayton}, G.~C., {Wolff}, M.~J., {Sofia}, U.~J., {Gordon}, K.~D., \& {Misselt}, K.~A. 2003, \apj, 588, 871, \dodoi{10.1086/374316}

\bibitem[{{Crutcher}(2012)}]{2012ARA&A..50...29C}
{Crutcher}, R.~M. 2012, \araa, 50, 29, \dodoi{10.1146/annurev-astro-081811-125514}

\bibitem[{{Crutcher} \& {Kemball}(2019)}]{2019CrutcherKemball}
{Crutcher}, R.~M., \& {Kemball}, A.~J. 2019, Frontiers in Astronomy and Space Sciences, 6, 66, \dodoi{10.3389/fspas.2019.00066}

\bibitem[{{Crutcher} {et~al.}(2004){Crutcher}, {Nutter}, {Ward-Thompson}, \& {Kirk}}]{crutcher04}
{Crutcher}, R.~M., {Nutter}, D.~J., {Ward-Thompson}, D., \& {Kirk}, J.~M. 2004, \apj, 600, 279, \dodoi{10.1086/379705}

\bibitem[{{Crutcher} {et~al.}(2010){Crutcher}, {Wandelt}, {Heiles}, {Falgarone}, \& {Troland}}]{crutcher2010}
{Crutcher}, R.~M., {Wandelt}, B., {Heiles}, C., {Falgarone}, E., \& {Troland}, T.~H. 2010, \apj, 725, 466, \dodoi{10.1088/0004-637X/725/1/466}

\bibitem[{{Dame} \& {Thaddeus}(2022)}]{2022ApJS..262....5D}
{Dame}, T.~M., \& {Thaddeus}, P. 2022, \apjs, 262, 5, \dodoi{10.3847/1538-4365/ac7e53}

\bibitem[{{Davis}(1951)}]{1951PhRv...81..890D}
{Davis}, L. 1951, \physrev, 81, 890, \dodoi{10.1103/PhysRev.81.890.2}

\bibitem[{{Dolginov} \& {Mitrofanov}(1976)}]{1976Ap&SS..43..291D}
{Dolginov}, A.~Z., \& {Mitrofanov}, I.~G. 1976, \apss, 43, 291, \dodoi{10.1007/BF00640010}

\bibitem[{{Fox} {et~al.}(2006){Fox}, {Savage}, {Wakker}, {Tripp}, \& {Sembach}}]{fox06}
{Fox}, A.~J., {Savage}, B.~D., {Wakker}, B.~P., {Tripp}, T.~M., \& {Sembach}, K.~R. 2006, in Astronomical Society of the Pacific Conference Series, Vol. 348, Astrophysics in the Far Ultraviolet: Five Years of Discovery with FUSE, ed. G.~{Sonneborn}, H.~W. {Moos}, \& B.~G. {Andersson}, 385

\bibitem[{{Gaia Collaboration} {et~al.}(2023){Gaia Collaboration}, {Vallenari}, {Brown}, {Prusti}, {de Bruijne}, {Arenou}, {Babusiaux}, {Biermann}, {Creevey}, {Ducourant}, {Evans}, {Eyer}, {Guerra}, {Hutton}, {Jordi}, {Klioner}, {Lammers}, {Lindegren}, {Luri}, {Mignard}, {Panem}, {Pourbaix}, {Randich}, {Sartoretti}, {Soubiran}, {Tanga}, {Walton}, {Bailer-Jones}, {Bastian}, {Drimmel}, {Jansen}, {Katz}, {Lattanzi}, {van Leeuwen}, {Bakker}, {Cacciari}, {Casta{\~n}eda}, {De Angeli}, {Fabricius}, {Fouesneau}, {Fr{\'e}mat}, {Galluccio}, {Guerrier}, {Heiter}, {Masana}, {Messineo}, {Mowlavi}, {Nicolas}, {Nienartowicz}, {Pailler}, {Panuzzo}, {Riclet}, {Roux}, {Seabroke}, {Sordo}, {Th{\'e}venin}, {Gracia-Abril}, {Portell}, {Teyssier}, {Altmann}, {Andrae}, {Audard}, {Bellas-Velidis}, {Benson}, {Berthier}, {Blomme}, {Burgess}, {Busonero}, {Busso}, {C{\'a}novas}, {Carry}, {Cellino}, {Cheek}, {Clementini}, {Damerdji}, {Davidson}, {de Teodoro}, {Nu{\~n}ez Campos}, {Delchambre}, {Dell'Oro}, {Esquej},
  {Fern{\'a}ndez-Hern{\'a}ndez}, {Fraile}, {Garabato}, {Garc{\'\i}a-Lario}, {Gosset}, {Haigron}, {Halbwachs}, {Hambly}, {Harrison}, {Hern{\'a}ndez}, {Hestroffer}, {Hodgkin}, {Holl}, {Jan{\ss}en}, {Jevardat de Fombelle}, {Jordan}, {Krone-Martins}, {Lanzafame}, {L{\"o}ffler}, {Marchal}, {Marrese}, {Moitinho}, {Muinonen}, {Osborne}, {Pancino}, {Pauwels}, {Recio-Blanco}, {Reyl{\'e}}, {Riello}, {Rimoldini}, {Roegiers}, {Rybizki}, {Sarro}, {Siopis}, {Smith}, {Sozzetti}, {Utrilla}, {van Leeuwen}, {Abbas}, {{\'A}brah{\'a}m}, {Abreu Aramburu}, {Aerts}, {Aguado}, {Ajaj}, {Aldea-Montero}, {Altavilla}, {{\'A}lvarez}, {Alves}, {Anders}, {Anderson}, {Anglada Varela}, {Antoja}, {Baines}, {Baker}, {Balaguer-N{\'u}{\~n}ez}, {Balbinot}, {Balog}, {Barache}, {Barbato}, {Barros}, {Barstow}, {Bartolom{\'e}}, {Bassilana}, {Bauchet}, {Becciani}, {Bellazzini}, {Berihuete}, {Bernet}, {Bertone}, {Bianchi}, {Binnenfeld}, {Blanco-Cuaresma}, {Blazere}, {Boch}, {Bombrun}, {Bossini}, {Bouquillon}, {Bragaglia}, {Bramante}, {Breedt},
  {Bressan}, {Brouillet}, {Brugaletta}, {Bucciarelli}, {Burlacu}, {Butkevich}, {Buzzi}, {Caffau}, {Cancelliere}, {Cantat-Gaudin}, {Carballo}, {Carlucci}, {Carnerero}, {Carrasco}, {Casamiquela}, {Castellani}, {Castro-Ginard}, {Chaoul}, {Charlot}, {Chemin}, {Chiaramida}, {Chiavassa}, {Chornay}, {Comoretto}, {Contursi}, {Cooper}, {Cornez}, {Cowell}, {Crifo}, {Cropper}, {Crosta}, {Crowley}, {Dafonte}, {Dapergolas}, {David}, {David}, {de Laverny}, {De Luise}, \& {De March}}]{gaiadr3}
{Gaia Collaboration}, {Vallenari}, A., {Brown}, A.~G.~A., {et~al.} 2023, \aap, 674, A1, \dodoi{10.1051/0004-6361/202243940}

\bibitem[{Heiles \& Troland(2005)}]{Heiles_2005}
Heiles, C., \& Troland, T.~H. 2005, The Astrophysical Journal, 624, 773, \dodoi{10.1086/428896}

\bibitem[{{Heyer} {et~al.}(2020){Heyer}, {Soler}, \& {Burkhart}}]{2020MNRAS.496.4546H}
{Heyer}, M., {Soler}, J.~D., \& {Burkhart}, B. 2020, \mnras, 496, 4546, \dodoi{10.1093/mnras/staa1760}

\bibitem[{{Hildebrand}(1983)}]{hildebrand1983}
{Hildebrand}, R.~H. 1983, \qjras, 24, 267

\bibitem[{{Hildebrand} {et~al.}(2009){Hildebrand}, {Kirby}, {Dotson}, {Houde}, \& {Vaillancourt}}]{2009Hildebrand}
{Hildebrand}, R.~H., {Kirby}, L., {Dotson}, J.~L., {Houde}, M., \& {Vaillancourt}, J.~E. 2009, \apj, 696, 567, \dodoi{10.1088/0004-637X/696/1/567}

\bibitem[{{Hoang} \& {Lazarian}(2008)}]{hoang08}
{Hoang}, T., \& {Lazarian}, A. 2008, \mnras, 388, 117, \dodoi{10.1111/j.1365-2966.2008.13249.x}

\bibitem[{{Hoang} \& {Lazarian}(2016)}]{thiem16}
---. 2016, \apj, 831, 159, \dodoi{10.3847/0004-637X/831/2/159}

\bibitem[{{Imara} \& {Burkhart}(2016)}]{2016ApJ...829..102I}
{Imara}, N., \& {Burkhart}, B. 2016, \apj, 829, 102, \dodoi{10.3847/0004-637X/829/2/102}

\bibitem[{{Jenkins}(2002)}]{jenkins2002}
{Jenkins}, E.~B. 2002, \apj, 580, 938, \dodoi{10.1086/343796}

\bibitem[{{Jo} {et~al.}(2017){Jo}, {Seon}, {Min}, {Edelstein}, \& {Han}}]{jo2017ApJS..231...21J}
{Jo}, Y.-S., {Seon}, K.-I., {Min}, K.-W., {Edelstein}, J., \& {Han}, W. 2017, \apjs, 231, 21, \dodoi{10.3847/1538-4365/aa8091}

\bibitem[{{Jones}(1989)}]{jones89}
{Jones}, T.~J. 1989, \apj, 346, 728, \dodoi{10.1086/168054}

\bibitem[{{Kwon} {et~al.}(2022){Kwon}, {Pattle}, {Sadavoy}, {Hull}, {Johnstone}, {Ward-Thompson}, {Di Francesco}, {Koch}, {Furuya}, {Doi}, {Le Gouellec}, {Hwang}, {Lyo}, {Soam}, {Tang}, {Hoang}, {Kirchschlager}, {Eswaraiah}, {Fanciullo}, {Kim}, {Onaka}, {K{\"o}nyves}, {Kang}, {Lee}, {Tamura}, {Bastien}, {Hasegawa}, {Lai}, {Qiu}, {Berry}, {Arzoumanian}, {Bourke}, {Byun}, {Chen}, {Chen}, {Chen}, {Chen}, {Ching}, {Cho}, {Choi}, {Choi}, {Chrysostomou}, {Chung}, {Coud{\'e}}, {Dai}, {Diep}, {Duan}, {Duan}, {Eden}, {Fiege}, {Fissel}, {Franzmann}, {Friberg}, {Friesen}, {Fuller}, {Gledhill}, {Graves}, {Greaves}, {Griffin}, {Gu}, {Han}, {Hatchell}, {Hayashi}, {Houde}, {Inoue}, {Inutsuka}, {Iwasaki}, {Jeong}, {Kang}, {Karoly}, {Kataoka}, {Kawabata}, {Kemper}, {Kim}, {Kim}, {Kim}, {Kim}, {Kim}, {Kirk}, {Kobayashi}, {Kusune}, {Kwon}, {Lacaille}, {Law}, {Lee}, {Lee}, {Lee}, {Lee}, {Lee}, {Li}, {Li}, {Li}, {Lin}, {Liu}, {Liu}, {Liu}, {Liu}, {Lu}, {Mairs}, {Matsumura}, {Matthews}, {Moriarty-Schieven}, {Nagata}, {Nakamura},
  {Nakanishi}, {Ngoc}, {Ohashi}, {Park}, {Parsons}, {Peretto}, {Priestley}, {Pyo}, {Qian}, {Rao}, {Rawlings}, {Rawlings}, {Retter}, {Richer}, {Rigby}, {Saito}, {Savini}, {Seta}, {Shimajiri}, {Shinnaga}, {Tahani}, {Tang}, {Tomisaka}, {Tram}, {Tsukamoto}, {Viti}, {Wang}, {Wang}, {Whitworth}, {Wu}, {Xie}, {Yen}, {Yoo}, {Yuan}, {Yun}, {Zenko}, {Zhang}, {Zhang}, {Zhang}, {Zhou}, {Zhu}, {de Looze}, {Andr{\'e}}, {Dowell}, {Eyres}, {Falle}, {Robitaille}, \& {Loo}}]{kwon2022}
{Kwon}, W., {Pattle}, K., {Sadavoy}, S., {et~al.} 2022, \apj, 926, 163, \dodoi{10.3847/1538-4357/ac4bbe}

\bibitem[{{Lazarian} \& {Hoang}(2008)}]{lazarian08}
{Lazarian}, A., \& {Hoang}, T. 2008, \apjl, 676, L25, \dodoi{10.1086/586706}

\bibitem[{{Le Gouellec} {et~al.}(2020){Le Gouellec}, {Maury}, {Guillet}, {Hull}, {Girart}, {Verliat}, {Mignon-Risse}, {Valdivia}, {Hennebelle}, {Gonz{\'a}lez}, \& {Louvet}}]{valentin2020}
{Le Gouellec}, V.~J.~M., {Maury}, A.~J., {Guillet}, V., {et~al.} 2020, \aap, 644, A11, \dodoi{10.1051/0004-6361/202038404}

\bibitem[{{Lee} {et~al.}(2012){Lee}, {Stanimirovi{\'c}}, {Douglas}, {Knee}, {Di Francesco}, {Gibson}, {Begum}, {Grcevich}, {Heiles}, {Korpela}, {Leroy}, {Peek}, {Pingel}, {Putman}, \& {Saul}}]{2012ApJ...748...75L}
{Lee}, M.-Y., {Stanimirovi{\'c}}, S., {Douglas}, K.~A., {et~al.} 2012, \apj, 748, 75, \dodoi{10.1088/0004-637X/748/2/75}

\bibitem[{{Magnani} {et~al.}(1985){Magnani}, {Blitz}, \& {Mundy}}]{magnani1985}
{Magnani}, L., {Blitz}, L., \& {Mundy}, L. 1985, \apj, 295, 402, \dodoi{10.1086/163385}

\bibitem[{{Mathis} {et~al.}(1977){Mathis}, {Rumpl}, \& {Nordsieck}}]{mrn77}
{Mathis}, J.~S., {Rumpl}, W., \& {Nordsieck}, K.~H. 1977, \apj, 217, 425, \dodoi{10.1086/155591}

\bibitem[{{McCammon} {et~al.}(1983){McCammon}, {Burrows}, {Sanders}, \& {Kraushaar}}]{mccammon1983}
{McCammon}, D., {Burrows}, D.~N., {Sanders}, W.~T., \& {Kraushaar}, W.~L. 1983, \apj, 269, 107, \dodoi{10.1086/161024}

\bibitem[{{McKee} \& {Ostriker}(2007)}]{2007ARA&A..45..565M}
{McKee}, C.~F., \& {Ostriker}, E.~C. 2007, \araa, 45, 565, \dodoi{10.1146/annurev.astro.45.051806.110602}

\bibitem[{{Mestel}(1966)}]{mestel1966}
{Mestel}, L. 1966, \mnras, 133, 265, \dodoi{10.1093/mnras/133.2.265}

\bibitem[{Mestel \& Spitzer(1956)}]{10.1093/mnras/116.5.503}
Mestel, L., \& Spitzer, L., J. 1956, Monthly Notices of the Royal Astronomical Society, 116, 503, \dodoi{10.1093/mnras/116.5.503}

\bibitem[{{Mouschovias}(1991)}]{1991ApJ...373..169M}
{Mouschovias}, T.~C. 1991, \apj, 373, 169, \dodoi{10.1086/170035}

\bibitem[{{Ostriker} {et~al.}(2001){Ostriker}, {Stone}, \& {Gammie}}]{2001ApJ...546..980O}
{Ostriker}, E.~C., {Stone}, J.~M., \& {Gammie}, C.~F. 2001, \apj, 546, 980, \dodoi{10.1086/318290}

\bibitem[{{Palmeirim} {et~al.}(2013){Palmeirim}, {Andr{\'e}}, {Kirk}, {Ward-Thompson}, {Arzoumanian}, {K{\"o}nyves}, {Didelon}, {Schneider}, {Benedettini}, {Bontemps}, {Di Francesco}, {Elia}, {Griffin}, {Hennemann}, {Hill}, {Martin}, {Men'shchikov}, {Molinari}, {Motte}, {Nguyen Luong}, {Nutter}, {Peretto}, {Pezzuto}, {Roy}, {Rygl}, {Spinoglio}, \& {White}}]{2013A&A...550A..38P}
{Palmeirim}, P., {Andr{\'e}}, P., {Kirk}, J., {et~al.} 2013, \aap, 550, A38, \dodoi{10.1051/0004-6361/201220500}

\bibitem[{{Pattle} {et~al.}(2023){Pattle}, {Fissel}, {Tahani}, {Liu}, \& {Ntormousi}}]{2022preStellarReview}
{Pattle}, K., {Fissel}, L., {Tahani}, M., {Liu}, T., \& {Ntormousi}, E. 2023, in Astronomical Society of the Pacific Conference Series, Vol. 534, Protostars and Planets VII, ed. S.~{Inutsuka}, Y.~{Aikawa}, T.~{Muto}, K.~{Tomida}, \& M.~{Tamura}, 193, \dodoi{10.48550/arXiv.2203.11179}

\bibitem[{{Pattle} {et~al.}(2019){Pattle}, {Lai}, {Hasegawa}, {Wang}, {Furuya}, {Ward-Thompson}, {Bastien}, {Coud{\'e}}, {Eswaraiah}, {Fanciullo}, {di Francesco}, {Hoang}, {Kim}, {Kwon}, {Lee}, {Liu}, {Liu}, {Matsumura}, {Onaka}, {Sadavoy}, \& {Soam}}]{katerice}
{Pattle}, K., {Lai}, S.-P., {Hasegawa}, T., {et~al.} 2019, \apj, 880, 27, \dodoi{10.3847/1538-4357/ab286f}

\bibitem[{{Peek} {et~al.}(2011){Peek}, {Heiles}, {Douglas}, {Lee}, {Grcevich}, {Stanimirovi{\'c}}, {Putman}, {Korpela}, {Gibson}, {Begum}, {Saul}, {Robishaw}, \& {Kr{\v{c}}o}}]{2011ApJS..194...20P}
{Peek}, J.~E.~G., {Heiles}, C., {Douglas}, K.~A., {et~al.} 2011, \apjs, 194, 20, \dodoi{10.1088/0067-0049/194/2/20}

\bibitem[{{Pingel} {et~al.}(2018){Pingel}, {Lee}, {Burkhart}, \& {Stanimirovi{\'c}}}]{2018ApJ...856..136P}
{Pingel}, N.~M., {Lee}, M.-Y., {Burkhart}, B., \& {Stanimirovi{\'c}}, S. 2018, \apj, 856, 136, \dodoi{10.3847/1538-4357/aab34b}

\bibitem[{{Planck Collaboration} {et~al.}(2015{\natexlab{a}}){Planck Collaboration}, {Ade}, {Aghanim}, {Alina}, {Alves}, {Armitage-Caplan}, {Arnaud}, {Arzoumanian}, {Ashdown}, {Atrio-Barandela}, {Aumont}, {Baccigalupi}, {Banday}, {Barreiro}, {Battaner}, {Benabed}, {Benoit-L{\'e}vy}, {Bernard}, {Bersanelli}, {Bielewicz}, {Bock}, {Bond}, {Borrill}, {Bouchet}, {Boulanger}, {Bracco}, {Burigana}, {Butler}, {Cardoso}, {Catalano}, {Chamballu}, {Chary}, {Chiang}, {Christensen}, {Colombi}, {Colombo}, {Combet}, {Couchot}, {Coulais}, {Crill}, {Curto}, {Cuttaia}, {Danese}, {Davies}, {Davis}, {de Bernardis}, {de Gouveia Dal Pino}, {de Rosa}, {de Zotti}, {Delabrouille}, {D{\'e}sert}, {Dickinson}, {Diego}, {Donzelli}, {Dor{\'e}}, {Douspis}, {Dunkley}, {Dupac}, {Efstathiou}, {En{\ss}lin}, {Eriksen}, {Falgarone}, {Ferri{\`e}re}, {Finelli}, {Forni}, {Frailis}, {Fraisse}, {Franceschi}, {Galeotta}, {Ganga}, {Ghosh}, {Giard}, {Giraud-H{\'e}raud}, {Gonz{\'a}lez-Nuevo}, {G{\'o}rski}, {Gregorio}, {Gruppuso}, {Guillet}, {Hansen},
  {Harrison}, {Helou}, {Hern{\'a}ndez-Monteagudo}, {Hildebrandt}, {Hivon}, {Hobson}, {Holmes}, {Hornstrup}, {Huffenberger}, {Jaffe}, {Jaffe}, {Jones}, {Juvela}, {Keih{\"a}nen}, {Keskitalo}, {Kisner}, {Kneissl}, {Knoche}, {Kunz}, {Kurki-Suonio}, {Lagache}, {L{\"a}hteenm{\"a}ki}, {Lamarre}, {Lasenby}, {Lawrence}, {Leahy}, {Leonardi}, {Levrier}, {Liguori}, {Lilje}, {Linden-V{\o}rnle}, {L{\'o}pez-Caniego}, {Lubin}, {Mac{\'\i}as-P{\'e}rez}, {Maffei}, {Magalh{\~a}es}, {Maino}, {Mandolesi}, {Maris}, {Marshall}, {Martin}, {Mart{\'\i}nez-Gonz{\'a}lez}, {Masi}, {Matarrese}, {Mazzotta}, {Melchiorri}, {Mendes}, {Mennella}, {Migliaccio}, {Miville-Desch{\^e}nes}, {Moneti}, {Montier}, {Morgante}, {Mortlock}, {Munshi}, {Murphy}, {Naselsky}, {Nati}, {Natoli}, {Netterfield}, {Noviello}, {Novikov}, {Novikov}, {Oxborrow}, {Pagano}, {Pajot}, {Paladini}, {Paoletti}, {Pasian}, {Pearson}, {Perdereau}, {Perotto}, {Perrotta}, {Piacentini}, {Piat}, {Pietrobon}, {Plaszczynski}, {Poidevin}, {Pointecouteau}, {Polenta}, {Popa}, {Pratt},
  {Prunet}, {Puget}, {Rachen}, {Reach}, {Rebolo}, {Reinecke}, {Remazeilles}, {Renault}, {Ricciardi}, {Riller}, {Ristorcelli}, {Rocha}, {Rosset}, {Roudier}, {Rubi{\~n}o-Mart{\'\i}n}, {Rusholme}, {Sandri}, {Savini}, {Scott}, {Spencer}, {Stolyarov}, {Stompor}, {Sudiwala}, {Sutton}, {Suur-Uski}, {Sygnet}, {Tauber}, {Terenzi}, {Toffolatti}, {Tomasi}, {Tristram}, {Tucci}, {Umana}, {Valenziano}, {Valiviita}, {Van Tent}, {Vielva}, {Villa}, {Wade}, {Wandelt}, {Zacchei}, \& {Zonca}}]{2015PlanckXIX}
{Planck Collaboration}, {Ade}, P.~A.~R., {Aghanim}, N., {et~al.} 2015{\natexlab{a}}, \aap, 576, A104, \dodoi{10.1051/0004-6361/201424082}

\bibitem[{{Planck Collaboration} {et~al.}(2015{\natexlab{b}}){Planck Collaboration}, {Ade}, {Aghanim}, {Alina}, {Alves}, {Aniano}, {Armitage-Caplan}, {Arnaud}, {Arzoumanian}, {Ashdown}, {Atrio-Barandela}, {Aumont}, {Baccigalupi}, {Banday}, {Barreiro}, {Battaner}, {Benabed}, {Benoit-L{\'e}vy}, {Bernard}, {Bersanelli}, {Bielewicz}, {Bond}, {Borrill}, {Bouchet}, {Boulanger}, {Bracco}, {Burigana}, {Cardoso}, {Catalano}, {Chamballu}, {Chiang}, {Christensen}, {Colombi}, {Colombo}, {Combet}, {Couchot}, {Coulais}, {Crill}, {Curto}, {Cuttaia}, {Danese}, {Davies}, {Davis}, {de Bernardis}, {de Rosa}, {de Zotti}, {Delabrouille}, {Dickinson}, {Diego}, {Donzelli}, {Dor{\'e}}, {Douspis}, {Dupac}, {Efstathiou}, {En{\ss}lin}, {Eriksen}, {Falgarone}, {Fanciullo}, {Ferri{\`e}re}, {Finelli}, {Forni}, {Frailis}, {Fraisse}, {Franceschi}, {Galeotta}, {Ganga}, {Ghosh}, {Giard}, {Giraud-H{\'e}raud}, {Gonz{\'a}lez-Nuevo}, {G{\'o}rski}, {Gregorio}, {Gruppuso}, {Guillet}, {Hansen}, {Harrison}, {Helou}, {Hern{\'a}ndez-Monteagudo},
  {Hildebrandt}, {Hivon}, {Hobson}, {Holmes}, {Hornstrup}, {Huffenberger}, {Jaffe}, {Jaffe}, {Jones}, {Juvela}, {Keih{\"a}nen}, {Keskitalo}, {Kisner}, {Kneissl}, {Knoche}, {Kunz}, {Kurki-Suonio}, {Lagache}, {Lamarre}, {Lasenby}, {Lawrence}, {Leonardi}, {Levrier}, {Liguori}, {Lilje}, {Linden-V{\o}rnle}, {L{\'o}pez-Caniego}, {Lubin}, {Mac{\'\i}as-P{\'e}rez}, {Maino}, {Mandolesi}, {Maris}, {Marshall}, {Martin}, {Mart{\'\i}nez-Gonz{\'a}lez}, {Masi}, {Matarrese}, {Mazzotta}, {Melchiorri}, {Mendes}, {Mennella}, {Migliaccio}, {Miville-Desch{\^e}nes}, {Moneti}, {Montier}, {Morgante}, {Mortlock}, {Munshi}, {Murphy}, {Naselsky}, {Nati}, {Natoli}, {Netterfield}, {Noviello}, {Novikov}, {Novikov}, {Oxborrow}, {Pagano}, {Pajot}, {Paoletti}, {Pasian}, {Pelkonen}, {Perdereau}, {Perotto}, {Perrotta}, {Piacentini}, {Piat}, {Pietrobon}, {Plaszczynski}, {Pointecouteau}, {Polenta}, {Popa}, {Pratt}, {Prunet}, {Puget}, {Rachen}, {Reinecke}, {Remazeilles}, {Renault}, {Ricciardi}, {Riller}, {Ristorcelli}, {Rocha}, {Rosset},
  {Roudier}, {Rusholme}, {Sandri}, {Scott}, {Soler}, {Spencer}, {Stolyarov}, {Stompor}, {Sudiwala}, {Sutton}, {Suur-Uski}, {Sygnet}, {Tauber}, {Terenzi}, {Toffolatti}, {Tomasi}, {Tristram}, {Tucci}, {Umana}, {Valenziano}, {Valiviita}, {Van Tent}, {Vielva}, {Villa}, {Wade}, {Wandelt}, \& {Zonca}}]{planckxx}
---. 2015{\natexlab{b}}, \aap, 576, A105, \dodoi{10.1051/0004-6361/201424086}

\bibitem[{{Planck Collaboration} {et~al.}(2016{\natexlab{a}}){Planck Collaboration}, {Adam}, {Ade}, {Alves}, {Ashdown}, {Aumont}, {Baccigalupi}, {Banday}, {Barreiro}, {Bartolo}, {Battaner}, {Benabed}, {Benoit-L{\'e}vy}, {Bernard}, {Bersanelli}, {Bielewicz}, {Bonavera}, {Bond}, {Borrill}, {Bouchet}, {Boulanger}, {Bucher}, {Burigana}, {Butler}, {Calabrese}, {Cardoso}, {Catalano}, {Chiang}, {Christensen}, {Colombo}, {Combet}, {Couchot}, {Crill}, {Curto}, {Cuttaia}, {Danese}, {Davis}, {de Bernardis}, {de Rosa}, {de Zotti}, {Delabrouille}, {Dickinson}, {Diego}, {Dolag}, {Dor{\'e}}, {Ducout}, {Dupac}, {Elsner}, {En{\ss}lin}, {Eriksen}, {Ferri{\`e}re}, {Finelli}, {Forni}, {Frailis}, {Fraisse}, {Franceschi}, {Galeotta}, {Ganga}, {Ghosh}, {Giard}, {Gjerl{\o}w}, {Gonz{\'a}lez-Nuevo}, {G{\'o}rski}, {Gregorio}, {Gruppuso}, {Gudmundsson}, {Hansen}, {Harrison}, {Hern{\'a}ndez-Monteagudo}, {Herranz}, {Hildebrandt}, {Hobson}, {Hornstrup}, {Hurier}, {Jaffe}, {Jaffe}, {Jones}, {Juvela}, {Keih{\"a}nen}, {Keskitalo}, {Kisner},
  {Knoche}, {Kunz}, {Kurki-Suonio}, {Lamarre}, {Lasenby}, {Lattanzi}, {Lawrence}, {Leahy}, {Leonardi}, {Levrier}, {Liguori}, {Lilje}, {Linden-V{\o}rnle}, {L{\'o}pez-Caniego}, {Lubin}, {Mac{\'\i}as-P{\'e}rez}, {Maggio}, {Maino}, {Mandolesi}, {Mangilli}, {Maris}, {Martin}, {Mart{\'\i}nez-Gonz{\'a}lez}, {Masi}, {Matarrese}, {Melchiorri}, {Mennella}, {Migliaccio}, {Miville-Desch{\^e}nes}, {Moneti}, {Montier}, {Morgante}, {Munshi}, {Murphy}, {Naselsky}, {Nati}, {Natoli}, {N{\o}rgaard-Nielsen}, {Oppermann}, {Orlando}, {Pagano}, {Pajot}, {Paladini}, {Paoletti}, {Pasian}, {Perotto}, {Pettorino}, {Piacentini}, {Piat}, {Pierpaoli}, {Plaszczynski}, {Pointecouteau}, {Polenta}, {Ponthieu}, {Pratt}, {Prunet}, {Puget}, {Rachen}, {Reinecke}, {Remazeilles}, {Renault}, {Renzi}, {Ristorcelli}, {Rocha}, {Rossetti}, {Roudier}, {Rubi{\~n}o-Mart{\'\i}n}, {Rusholme}, {Sandri}, {Santos}, {Savelainen}, {Scott}, {Spencer}, {Stolyarov}, {Stompor}, {Strong}, {Sudiwala}, {Sunyaev}, {Suur-Uski}, {Sygnet}, {Tauber}, {Terenzi}, {Toffolatti},
  {Tomasi}, {Tristram}, {Tucci}, {Valenziano}, {Valiviita}, {Van Tent}, {Vielva}, {Villa}, {Wade}, {Wandelt}, {Wehus}, {Yvon}, {Zacchei}, \& {Zonca}}]{2016A&A...596A.103P}
{Planck Collaboration}, {Adam}, R., {Ade}, P.~A.~R., {et~al.} 2016{\natexlab{a}}, \aap, 596, A103, \dodoi{10.1051/0004-6361/201528033}

\bibitem[{{Planck Collaboration} {et~al.}(2016{\natexlab{b}}){Planck Collaboration}, {Ade}, {Aghanim}, {Alves}, {Arnaud}, {Arzoumanian}, {Ashdown}, {Aumont}, {Baccigalupi}, {Banday}, {Barreiro}, {Bartolo}, {Battaner}, {Benabed}, {Beno{\^\i}t}, {Benoit-L{\'e}vy}, {Bernard}, {Bersanelli}, {Bielewicz}, {Bock}, {Bonavera}, {Bond}, {Borrill}, {Bouchet}, {Boulanger}, {Bracco}, {Burigana}, {Calabrese}, {Cardoso}, {Catalano}, {Chiang}, {Christensen}, {Colombo}, {Combet}, {Couchot}, {Crill}, {Curto}, {Cuttaia}, {Danese}, {Davies}, {Davis}, {de Bernardis}, {de Rosa}, {de Zotti}, {Delabrouille}, {Dickinson}, {Diego}, {Dole}, {Donzelli}, {Dor{\'e}}, {Douspis}, {Ducout}, {Dupac}, {Efstathiou}, {Elsner}, {En{\ss}lin}, {Eriksen}, {Falceta-Gon{\c{c}}alves}, {Falgarone}, {Ferri{\`e}re}, {Finelli}, {Forni}, {Frailis}, {Fraisse}, {Franceschi}, {Frejsel}, {Galeotta}, {Galli}, {Ganga}, {Ghosh}, {Giard}, {Gjerl{\o}w}, {Gonz{\'a}lez-Nuevo}, {G{\'o}rski}, {Gregorio}, {Gruppuso}, {Gudmundsson}, {Guillet}, {Harrison}, {Helou},
  {Hennebelle}, {Henrot-Versill{\'e}}, {Hern{\'a}ndez-Monteagudo}, {Herranz}, {Hildebrandt}, {Hivon}, {Holmes}, {Hornstrup}, {Huffenberger}, {Hurier}, {Jaffe}, {Jaffe}, {Jones}, {Juvela}, {Keih{\"a}nen}, {Keskitalo}, {Kisner}, {Knoche}, {Kunz}, {Kurki-Suonio}, {Lagache}, {Lamarre}, {Lasenby}, {Lattanzi}, {Lawrence}, {Leonardi}, {Levrier}, {Liguori}, {Lilje}, {Linden-V{\o}rnle}, {L{\'o}pez-Caniego}, {Lubin}, {Mac{\'\i}as-P{\'e}rez}, {Maino}, {Mandolesi}, {Mangilli}, {Maris}, {Martin}, {Mart{\'\i}nez-Gonz{\'a}lez}, {Masi}, {Matarrese}, {Melchiorri}, {Mendes}, {Mennella}, {Migliaccio}, {Miville-Desch{\^e}nes}, {Moneti}, {Montier}, {Morgante}, {Mortlock}, {Munshi}, {Murphy}, {Naselsky}, {Nati}, {Netterfield}, {Noviello}, {Novikov}, {Novikov}, {Oppermann}, {Oxborrow}, {Pagano}, {Pajot}, {Paladini}, {Paoletti}, {Pasian}, {Perotto}, {Pettorino}, {Piacentini}, {Piat}, {Pierpaoli}, {Pietrobon}, {Plaszczynski}, {Pointecouteau}, {Polenta}, {Ponthieu}, {Pratt}, {Prunet}, {Puget}, {Rachen}, {Reinecke}, {Remazeilles},
  {Renault}, {Renzi}, {Ristorcelli}, {Rocha}, {Rossetti}, {Roudier}, {Rubi{\~n}o-Mart{\'\i}n}, {Rusholme}, {Sandri}, {Santos}, {Savelainen}, {Savini}, {Scott}, {Soler}, {Stolyarov}, {Sudiwala}, {Sutton}, {Suur-Uski}, {Sygnet}, {Tauber}, {Terenzi}, {Toffolatti}, {Tomasi}, {Tristram}, {Tucci}, {Umana}, {Valenziano}, {Valiviita}, {Van Tent}, {Vielva}, {Villa}, {Wade}, {Wandelt}, {Wehus}, {Ysard}, {Yvon}, \& {Zonca}}]{Planck2016_xxxv}
{Planck Collaboration}, {Ade}, P.~A.~R., {Aghanim}, N., {et~al.} 2016{\natexlab{b}}, \aap, 586, A138, \dodoi{10.1051/0004-6361/201525896}

\bibitem[{{Planck Collaboration} {et~al.}(2016{\natexlab{c}}){Planck Collaboration}, {Ade}, {Aghanim}, {Arnaud}, {Ashdown}, {Aumont}, {Baccigalupi}, {Banday}, {Barreiro}, {Bartolo}, {Battaner}, {Benabed}, {Beno{\^\i}t}, {Benoit-L{\'e}vy}, {Bernard}, {Bersanelli}, {Bielewicz}, {Bonaldi}, {Bonavera}, {Bond}, {Borrill}, {Bouchet}, {Boulanger}, {Bucher}, {Burigana}, {Butler}, {Calabrese}, {Catalano}, {Chamballu}, {Chiang}, {Christensen}, {Clements}, {Colombi}, {Colombo}, {Combet}, {Couchot}, {Coulais}, {Crill}, {Curto}, {Cuttaia}, {Danese}, {Davies}, {Davis}, {de Bernardis}, {de Rosa}, {de Zotti}, {Delabrouille}, {D{\'e}sert}, {Dickinson}, {Diego}, {Dole}, {Donzelli}, {Dor{\'e}}, {Douspis}, {Ducout}, {Dupac}, {Efstathiou}, {Elsner}, {En{\ss}lin}, {Eriksen}, {Falgarone}, {Fergusson}, {Finelli}, {Forni}, {Frailis}, {Fraisse}, {Franceschi}, {Frejsel}, {Galeotta}, {Galli}, {Ganga}, {Giard}, {Giraud-H{\'e}raud}, {Gjerl{\o}w}, {Gonz{\'a}lez-Nuevo}, {G{\'o}rski}, {Gratton}, {Gregorio}, {Gruppuso}, {Gudmundsson},
  {Hansen}, {Hanson}, {Harrison}, {Helou}, {Henrot-Versill{\'e}}, {Hern{\'a}ndez-Monteagudo}, {Herranz}, {Hildebrandt}, {Hivon}, {Hobson}, {Holmes}, {Hornstrup}, {Hovest}, {Huffenberger}, {Hurier}, {Jaffe}, {Jaffe}, {Jones}, {Juvela}, {Keih{\"a}nen}, {Keskitalo}, {Kisner}, {Knoche}, {Kunz}, {Kurki-Suonio}, {Lagache}, {Lamarre}, {Lasenby}, {Lattanzi}, {Lawrence}, {Leonardi}, {Lesgourgues}, {Levrier}, {Liguori}, {Lilje}, {Linden-V{\o}rnle}, {L{\'o}pez-Caniego}, {Lubin}, {Mac{\'\i}as-P{\'e}rez}, {Maggio}, {Maino}, {Mandolesi}, {Mangilli}, {Marshall}, {Martin}, {Mart{\'\i}nez-Gonz{\'a}lez}, {Masi}, {Matarrese}, {Mazzotta}, {McGehee}, {Melchiorri}, {Mendes}, {Mennella}, {Migliaccio}, {Mitra}, {Miville-Desch{\^e}nes}, {Moneti}, {Montier}, {Morgante}, {Mortlock}, {Moss}, {Munshi}, {Murphy}, {Naselsky}, {Nati}, {Natoli}, {Netterfield}, {N{\o}rgaard-Nielsen}, {Noviello}, {Novikov}, {Novikov}, {Oxborrow}, {Paci}, {Pagano}, {Pajot}, {Paladini}, {Paoletti}, {Pasian}, {Patanchon}, {Pearson}, {Pelkonen}, {Perdereau},
  {Perotto}, {Perrotta}, {Pettorino}, {Piacentini}, {Piat}, {Pierpaoli}, {Pietrobon}, {Plaszczynski}, {Pointecouteau}, {Polenta}, {Pratt}, {Pr{\'e}zeau}, {Prunet}, {Puget}, {Rachen}, {Reach}, {Rebolo}, {Reinecke}, {Remazeilles}, {Renault}, {Renzi}, {Ristorcelli}, {Rocha}, {Rosset}, {Rossetti}, {Roudier}, {Rubi{\~n}o-Mart{\'\i}n}, {Rusholme}, {Sandri}, {Santos}, {Savelainen}, {Savini}, {Scott}, {Seiffert}, {Shellard}, {Spencer}, {Stolyarov}, \& {Sudiwala}}]{planck_pgcc}
---. 2016{\natexlab{c}}, \aap, 594, A28, \dodoi{10.1051/0004-6361/201525819}

\bibitem[{{Planck Collaboration} {et~al.}(2016{\natexlab{d}}){Planck Collaboration}, {Adam}, {Ade}, {Aghanim}, {Alves}, {Arnaud}, {Arzoumanian}, {Ashdown}, {Aumont}, {Baccigalupi}, {Banday}, {Barreiro}, {Bartolo}, {Battaner}, {Benabed}, {Benoit-L{\'e}vy}, {Bernard}, {Bersanelli}, {Bielewicz}, {Bonaldi}, {Bonavera}, {Bond}, {Borrill}, {Bouchet}, {Boulanger}, {Bracco}, {Burigana}, {Butler}, {Calabrese}, {Cardoso}, {Catalano}, {Chamballu}, {Chiang}, {Christensen}, {Colombi}, {Colombo}, {Combet}, {Couchot}, {Crill}, {Curto}, {Cuttaia}, {Danese}, {Davies}, {Davis}, {de Bernardis}, {de Rosa}, {de Zotti}, {Delabrouille}, {Dickinson}, {Diego}, {Dole}, {Donzelli}, {Dor{\'e}}, {Douspis}, {Ducout}, {Dupac}, {Efstathiou}, {Elsner}, {En{\ss}lin}, {Eriksen}, {Falgarone}, {Ferri{\`e}re}, {Finelli}, {Forni}, {Frailis}, {Fraisse}, {Franceschi}, {Frejsel}, {Galeotta}, {Galli}, {Ganga}, {Ghosh}, {Giard}, {Gjerl{\o}w}, {Gonz{\'a}lez-Nuevo}, {G{\'o}rski}, {Gregorio}, {Gruppuso}, {Guillet}, {Hansen}, {Hanson}, {Harrison},
  {Henrot-Versill{\'e}}, {Hern{\'a}ndez-Monteagudo}, {Herranz}, {Hildebrandt}, {Hivon}, {Hobson}, {Holmes}, {Hovest}, {Huffenberger}, {Hurier}, {Jaffe}, {Jaffe}, {Jones}, {Juvela}, {Keih{\"a}nen}, {Keskitalo}, {Kisner}, {Kneissl}, {Knoche}, {Kunz}, {Kurki-Suonio}, {Lagache}, {Lamarre}, {Lasenby}, {Lattanzi}, {Lawrence}, {Leonardi}, {Levrier}, {Liguori}, {Lilje}, {Linden-V{\o}rnle}, {L{\'o}pez-Caniego}, {Lubin}, {Mac{\'\i}as-P{\'e}rez}, {Maffei}, {Maino}, {Mandolesi}, {Maris}, {Marshall}, {Martin}, {Mart{\'\i}nez-Gonz{\'a}lez}, {Masi}, {Matarrese}, {Mazzotta}, {Melchiorri}, {Mendes}, {Mennella}, {Migliaccio}, {Miville-Desch{\^e}nes}, {Moneti}, {Montier}, {Morgante}, {Mortlock}, {Munshi}, {Murphy}, {Naselsky}, {Natoli}, {N{\o}rgaard-Nielsen}, {Noviello}, {Novikov}, {Novikov}, {Oppermann}, {Oxborrow}, {Pagano}, {Pajot}, {Paoletti}, {Pasian}, {Perdereau}, {Perotto}, {Perrotta}, {Pettorino}, {Piacentini}, {Piat}, {Plaszczynski}, {Pointecouteau}, {Polenta}, {Ponthieu}, {Popa}, {Pratt}, {Prunet}, {Puget}, {Rachen},
  {Reach}, {Reinecke}, {Remazeilles}, {Renault}, {Ristorcelli}, {Rocha}, {Roudier}, {Rubi{\~n}o-Mart{\'\i}n}, {Rusholme}, {Sandri}, {Santos}, {Savini}, {Scott}, {Soler}, {Spencer}, {Stolyarov}, {Sudiwala}, {Sunyaev}, {Sutton}, {Suur-Uski}, {Sygnet}, {Tauber}, {Terenzi}, {Toffolatti}, {Tomasi}, {Tristram}, {Tucci}, {Umana}, {Valenziano}, {Valiviita}, {Van Tent}, {Vielva}, {Villa}, {Wade}, {Wandelt}, \& {Wehus}}]{Planck2016_xxxii}
{Planck Collaboration}, {Adam}, R., {Ade}, P.~A.~R., {et~al.} 2016{\natexlab{d}}, \aap, 586, A135, \dodoi{10.1051/0004-6361/201425044}

\bibitem[{{Planck Collaboration} {et~al.}(2020){Planck Collaboration}, {Aghanim}, {Akrami}, {Alves}, {Ashdown}, {Aumont}, {Baccigalupi}, {Ballardini}, {Banday}, {Barreiro}, {Bartolo}, {Basak}, {Benabed}, {Bernard}, {Bersanelli}, {Bielewicz}, {Bock}, {Bond}, {Borrill}, {Bouchet}, {Boulanger}, {Bracco}, {Bucher}, {Burigana}, {Calabrese}, {Cardoso}, {Carron}, {Chary}, {Chiang}, {Colombo}, {Combet}, {Crill}, {Cuttaia}, {de Bernardis}, {de Zotti}, {Delabrouille}, {Delouis}, {Di Valentino}, {Dickinson}, {Diego}, {Dor{\'e}}, {Douspis}, {Ducout}, {Dupac}, {Efstathiou}, {Elsner}, {En{\ss}lin}, {Eriksen}, {Falgarone}, {Fantaye}, {Fernandez-Cobos}, {Ferri{\`e}re}, {Finelli}, {Forastieri}, {Frailis}, {Fraisse}, {Franceschi}, {Frolov}, {Galeotta}, {Galli}, {Ganga}, {G{\'e}nova-Santos}, {Gerbino}, {Ghosh}, {Gonz{\'a}lez-Nuevo}, {G{\'o}rski}, {Gratton}, {Green}, {Gruppuso}, {Gudmundsson}, {Guillet}, {Handley}, {Hansen}, {Helou}, {Herranz}, {Hivon}, {Huang}, {Jaffe}, {Jones}, {Keih{\"a}nen}, {Keskitalo}, {Kiiveri}, {Kim},
  {Krachmalnicoff}, {Kunz}, {Kurki-Suonio}, {Lagache}, {Lamarre}, {Lasenby}, {Lattanzi}, {Lawrence}, {Le Jeune}, {Levrier}, {Liguori}, {Lilje}, {Lindholm}, {L{\'o}pez-Caniego}, {Lubin}, {Ma}, {Mac{\'\i}as-P{\'e}rez}, {Maggio}, {Maino}, {Mandolesi}, {Mangilli}, {Marcos-Caballero}, {Maris}, {Martin}, {Mart{\'\i}nez-Gonz{\'a}lez}, {Matarrese}, {Mauri}, {McEwen}, {Melchiorri}, {Mennella}, {Migliaccio}, {Miville-Desch{\^e}nes}, {Molinari}, {Moneti}, {Montier}, {Morgante}, {Moss}, {Natoli}, {Pagano}, {Paoletti}, {Patanchon}, {Perrotta}, {Pettorino}, {Piacentini}, {Polastri}, {Polenta}, {Puget}, {Rachen}, {Reinecke}, {Remazeilles}, {Renzi}, {Ristorcelli}, {Rocha}, {Rosset}, {Roudier}, {Rubi{\~n}o-Mart{\'\i}n}, {Ruiz-Granados}, {Salvati}, {Sandri}, {Savelainen}, {Scott}, {Sirignano}, {Sunyaev}, {Suur-Uski}, {Tauber}, {Tavagnacco}, {Tenti}, {Toffolatti}, {Tomasi}, {Trombetti}, {Valiviita}, {Vansyngel}, {Van Tent}, {Vielva}, {Villa}, {Vittorio}, {Wandelt}, {Wehus}, {Zacchei}, \& {Zonca}}]{planckxii}
{Planck Collaboration}, {Aghanim}, N., {Akrami}, Y., {et~al.} 2020, \aap, 641, A12, \dodoi{10.1051/0004-6361/201833885}

\bibitem[{{Saxena} {et~al.}(2025){Saxena}, {Haworth}, {Burkhart}, {Dharmawardena}, {Gillen}, {Pattle}, {Karoly}, \& {Hamden}}]{saxena2025}
{Saxena}, S., {Haworth}, T.~J., {Burkhart}, B., {et~al.} 2025, arXiv e-prints, arXiv:2504.17850.
\newblock \doarXiv{2504.17850}

\bibitem[{{Shore} {et~al.}(2003){Shore}, {Magnani}, {LaRosa}, \& {McCarthy}}]{shore03}
{Shore}, S.~N., {Magnani}, L., {LaRosa}, T.~N., \& {McCarthy}, M.~N. 2003, \apj, 593, 413, \dodoi{10.1086/376355}

\bibitem[{{Skalidis} \& {Tassis}(2021)}]{2021Skalidis}
{Skalidis}, R., \& {Tassis}, K. 2021, \aap, 647, A186, \dodoi{10.1051/0004-6361/202039779}

\bibitem[{{Smith} \& {Cox}(2001)}]{smith2001}
{Smith}, R.~K., \& {Cox}, D.~P. 2001, \apjs, 134, 283, \dodoi{10.1086/320850}

\bibitem[{{Snowden} {et~al.}(1998){Snowden}, {Egger}, {Finkbeiner}, {Freyberg}, \& {Plucinsky}}]{snowden1998}
{Snowden}, S.~L., {Egger}, R., {Finkbeiner}, D.~P., {Freyberg}, M.~J., \& {Plucinsky}, P.~P. 1998, \apj, 493, 715, \dodoi{10.1086/305135}

\bibitem[{{Soler} {et~al.}(2018){Soler}, {Bracco}, \& {Pon}}]{soler2018}
{Soler}, J.~D., {Bracco}, A., \& {Pon}, A. 2018, \aap, 609, L3, \dodoi{10.1051/0004-6361/201732203}

\bibitem[{{Soler} \& {Hennebelle}(2017)}]{2017A&A...607A...2S}
{Soler}, J.~D., \& {Hennebelle}, P. 2017, \aap, 607, A2, \dodoi{10.1051/0004-6361/201731049}

\bibitem[{{Soler} {et~al.}(2013){Soler}, {Hennebelle}, {Martin}, {Miville-Desch{\^e}nes}, {Netterfield}, \& {Fissel}}]{2013ApJ...774..128S}
{Soler}, J.~D., {Hennebelle}, P., {Martin}, P.~G., {et~al.} 2013, \apj, 774, 128, \dodoi{10.1088/0004-637X/774/2/128}

\bibitem[{{Soler} {et~al.}(2017){Soler}, {Ade}, {Angil{\`e}}, {Ashton}, {Benton}, {Devlin}, {Dober}, {Fissel}, {Fukui}, {Galitzki}, {Gandilo}, {Hennebelle}, {Klein}, {Li}, {Korotkov}, {Martin}, {Matthews}, {Moncelsi}, {Netterfield}, {Novak}, {Pascale}, {Poidevin}, {Santos}, {Savini}, {Scott}, {Shariff}, {Thomas}, {Tucker}, {Tucker}, \& {Ward-Thompson}}]{soler2017}
{Soler}, J.~D., {Ade}, P.~A.~R., {Angil{\`e}}, F.~E., {et~al.} 2017, \aap, 603, A64, \dodoi{10.1051/0004-6361/201730608}

\bibitem[{{Sternberg} {et~al.}(2021){Sternberg}, {Gurman}, \& {Bialy}}]{Sternberg_2021}
{Sternberg}, A., {Gurman}, A., \& {Bialy}, S. 2021, \apj, 920, 83, \dodoi{10.3847/1538-4357/ac167b}

\bibitem[{{Sun} {et~al.}(2021){Sun}, {Jiang}, {Zhao}, \& {Ren}}]{sun2021}
{Sun}, M., {Jiang}, B., {Zhao}, H., \& {Ren}, Y. 2021, \apjs, 256, 46, \dodoi{10.3847/1538-4365/ac1601}

\bibitem[{{Verschuur} \& {Magnani}(1994)}]{verschuur1994}
{Verschuur}, G.~L., \& {Magnani}, L. 1994, \aj, 107, 287, \dodoi{10.1086/116853}

\bibitem[{Virtanen {et~al.}(2020)Virtanen, Gommers, Oliphant, Haberland, Reddy, Cournapeau, Burovski, Peterson, Weckesser, Bright, {van der Walt}, Brett, Wilson, Millman, Mayorov, Nelson, Jones, Kern, Larson, Carey, Polat, Feng, Moore, {VanderPlas}, Laxalde, Perktold, Cimrman, Henriksen, Quintero, Harris, Archibald, Ribeiro, Pedregosa, {van Mulbregt}, \& {SciPy 1.0 Contributors}}]{2020SciPy-NMeth}
Virtanen, P., Gommers, R., Oliphant, T.~E., {et~al.} 2020, Nature Methods, 17, 261, \dodoi{10.1038/s41592-019-0686-2}

\bibitem[{{Zucker} {et~al.}(2022){Zucker}, {Goodman}, {Alves}, {Bialy}, {Foley}, {Speagle}, {Gro{\^I}{\texttwosuperior}schedl}, {Finkbeiner}, {Burkert}, {Khimey}, \& {Swiggum}}]{zucker22}
{Zucker}, C., {Goodman}, A.~A., {Alves}, J., {et~al.} 2022, \nat, 601, 334, \dodoi{10.1038/s41586-021-04286-5}

\end{thebibliography}
\bibliographystyle{aasjournal}

\end{document}